# 2 Google Scholar as a data source for research assessment

Emilio Delgado López-Cózar
Facultad de Comunicación y Documentación, Universidad de Granada (Spain)
E-mail: edelgado@ugr.es

Enrique Orduna-Malea
Universitat Politècnica de València (Spain)
E-mail: enorma@upv.es

Alberto Martín-Martín
Facultad de Comunicación y Documentación, Universidad de Granada (Spain)
E-mail: albertomartin@ugr.es

**Abstract**:

The launch of Google Scholar (GS) marked the beginning of a revolution in the scientific information market. This search engine, unlike traditional databases, automatically indexes information from the academic web. Its ease of use, together with its wide coverage and fast indexing speed, have made it the first tool most scientists currently turn to when they need to carry out a literature search. Additionally, the fact that its search results were accompanied from the beginning by citation counts, as well as the later development of secondary products which leverage this citation data (such as Google Scholar Metrics and Google Scholar Citations), made many scientists wonder about its potential as a source of data for bibliometric analyses. The goal of this chapter is to lay the foundations for the use of GS as a supplementary source (and in some disciplines, arguably the best alternative) for scientific evaluation. First, we present a general overview of how GS works. Second, we present empirical evidences about its main characteristics (size, coverage, and growth rate). Third, we carry out a systematic analysis of the main limitations this search engine presents as a tool for the evaluation of scientific performance. Lastly, we discuss the main differences between GS and other more traditional bibliographic databases in light of the correlations found between their citation data. We conclude that Google Scholar presents a broader view of the academic world because it has brought to light a great amount of sources that were not previously visible.

## 2.1 Introduction

The development of the field of Bibliometrics has always been reliant on the availability of large-scale sources of metadata about scientific publications, which are ultimately the raw materials used by bibliometricians to carry out their analyses [1]. The creation of the first citation indexes by Eugene Garfield (Science Citation Index in 1964, Social Science Citation Index in 1973, and Arts & Humanities Citation Index in 1978) turned out to be a crucial turning point that enabled the development of modern bibliometric studies. His novel approach to bibliographic information systems opened the way to a completely new way of assessing scientific performance [2].

By indexing not only the articles published in scientific journals, but also the bibliographic references included in these articles, it was possible for the first time to track the relationships between scientists, journals, and institutions through the main tangible outputs these entities produce: scientific documents. Thus, the databases of the Institute for Scientific Information (now part of Clarivate's Web of Science) became the first, and for a long time the only available sources of data for bibliometric analyses, exercising an almost absolute monopoly in this field. The use of other specialized databases (such as Medline, Chemical Abstracts, Inspec, or Biosis) for bibliometric purposes was testimonial, since they did not offer citation data nor other fields (e.g. full affiliations of the authors) which are vital to produce useful bibliometric reports.

It wasn't until the first decade of the 21$^{st}$ century that this monopoly was seriously challenged. Elsevier launched its Scopus citation database on the 3$^{rd}$ of November of 2004. Just a few weeks later (18$^{th}$ of November) Google Scholar (GS) was also launched. Scopus was conceived as a traditional subscription-based bibliographic database (which indexed a specific set of journals and conference proceedings) and was clearly a direct competitor to Web of Science (WoS). GS departed entirely from this approach, following instead the path of its big brother, the Google search engine, a decision that greatly impacted its design and coverage.

Simply put, GS is a specialised search engine that only indexes academic documents [3-4]. Google Scholar's spiders constantly crawl the websites of universities, scientific publishers, topic and institutional repositories, databases, aggregators, library catalogues, and any other web spaces where they might find academic-like materials, regardless of their subject or language. GS indexes documents from the whole range of academic document types (books, book chapters, journal and conference articles, teaching materials, theses, posters, presentations, reports, patents, etc.). Unlike the cumulative and selective nature of WoS and Scopus, GS is dynamic: it reflects the state of the web as it is visible to their search robots and to the majority of users at a specific moment in time. Documents that for any reason become unavailable on the Web will eventually disappear from GS too, as will the citations they provided to other documents [5].

GS, like the Google search engine before it, achieved instant success among users worldwide. The reason is easy to understand: GS finds most of the scientific information that circulates around the web in an easy and fast manner. Perhaps most importantly, it is free, unlike most of the bibliographic databases that existed before it, which are often only accessible through costly national or university-level subscriptions.

Google Scholar is currently the tool most users first turn to when they need to carry out a literature search. This has been evidenced by numerous studies [6-11]. Bosman and Kramer's study is the most recent and large-scale study on the matter. They conducted a survey on the changing landscape of scholarly communication between May 2015 and February 2016, obtaining more than 20,000 responses from researchers, university students, librarians, and other members of the scholarly community. To the question *What tools do you use to search literature?* GS emerged as the preferred option, selected by 89% of the respondents, followed at a great distance by WoS (41%), Pubmed (40%), Others (36%), and Scopus (26%).

Since its launch in 2004, Google Scholar's interface has gone through several renovations, but the really important changes (updates to its algorithms, coverage) usually happened under the hood, unbeknownst to most users. Some developments, however, didn't go unnoticed. We are referring, of course, to the creation of its two secondary products: Google Scholar Citations (GSC) and Google Scholar Metrics (GSM). GSC was launched in July 2011 and provided a platform in which users could easily create an academic profile by pulling their publications from the data available in GS. Most interestingly, these profiles also displayed several author-level bibliometric indicators [12]. GSM was born on April 2012 as a ranking of scholarly publications according to their h-index calculated from Google Scholar data. This tool provides an easy way to identify the most influential publications (journals, proceedings, and repositories) and articles published in recent years [13].

Although these two tools never lose sight of Google Scholar's main purpose (they are intended to serve as search tools, one to find relevant researchers, the other to find influential articles and publications), they use bibliometric indicators as an evidence of relevance. For the first time, the GS team decided to put the citation data available in GS to other uses. Until the creation of those products, citation counts were only used as one of the parameters to rank documents in a search, and a search aid for users (*Cited by* links in GS).

The availability of citation data in GS and its secondary products GSC and GSM has attracted the attention of some bibliometricians, and even scientists from other fields, who have realized that the data available in GS provides a much more comprehensive insight into the impact publications have on their respective academic communities than the data available in other citation databases. However, the use of GS for bibliometric purposes was never one of the applications GS's developers intended for this product, and so an exhaustive critical evaluation that analyzes its suitability for bibliometric analyses is necessary.

In order to do this, this chapter first presents a general overview of how GS and its secondary products GSC and GSM work, their inclusion policies, and how they respond (results offered) when specific stimuli (user queries) are applied. Second, we present empirical evidences regarding its size, evolution (growth rate, indexing speed), coverage (publishers, repositories, bibliographic databases, catalogues), and diversity (subjects, languages, document types). Third, we carry out a systematic analysis of the main limitations this search engine presents as a tool for the evaluation of scientific performance. Lastly, we discuss the main differences between Google Scholar and other traditional bibliographic databases in light of the correlations found between their citation data at the level of authors, documents, and journals.

**2.2 Basic functioning of Google Scholar**

In this section we first present a concise but accurate description of how the GS search engine works. Secondly, we describe its main inclusion criteria (both for sources and, especially, for documents). Lastly, we will briefly outline Google Scholar Citations (GSC) and Google Scholar Metrics (GSM).

**2.2.1 The Academic Search Engine**

Classic bibliographic databases usually work on the principle of whitelists. They first generate a whitelist of sources which meet some specific criteria (quality, subject scope…), and then index all the publications that appear in these sources. The historical tendency to select some specific sources (mainly journals) and not other channels for the dissemination of academic results (conference proceedings, books, reports, etc.) responds mainly to two reasons. First, it is a question of efficiency, usually referred to as Bradford's law of scattering [14], thanks to which we know that for any given topic, a small core of journals provides most of the articles on that topic. When faced with technological and economic constraints, maximizing returns by selecting only the core of journals that will be most useful for a given purpose seems a logical and natural response. The other reason has to do with the evaluative use of these databases. Due to their visibility and prestige, most authors want to publish their articles in these core journals, increasing the competition to get a manuscript accepted in these journals. The limited space for publication of the printed era, as well as the higher standards to which articles are held in these journals are what helps project their image of prestige. To publish an article in a core journal is a difficult task, something that only the best researchers manage to do. In the same way, receiving a citation from an article published in a core journal also lends prestige to the cited article and its authors. This is the road to research excellence.

It goes without saying that this traditional approach (which prioritises the optimization of resources and excellence) is not without its merits, and has played an important role up until now. However, the irruption of GS represents a break from this paradigm. Unlike traditional bibliographic databases, which are selective by nature, GS parses the entire academic web, indexing every scholarly document it finds regardless of its quality, and doesn't differentiate between peer-reviewed and non-peer-reviewed content. GS is, rather, an academic search engine [3] with a bibliographic database that grows in a (mostly) unsupervised manner, and which has one clear purpose: facilitating the discovery of academic literature to everyone worldwide.

This unsupervised indexing process is possible thanks to the automated bots (sometimes also called spiders) that GS deploys throughout the web, similarly to how the Google search engine crawls the web as well. In GS, these bots are trained to locate academic resources, index their full texts (whenever possible), and extract their bibliographic descriptions (metadata). The process ends with the automated creation of a bibliographic record that is ready to be included in a search engine results page (SERP) when it is deemed relevant for a particular query. In order for a particular academic website to be successfully indexed in GS, certain technical requirements must be met, i.e. bots must be allowed to enter the website, there must be an easy-to-follow route to the article pages, and certain metadata must be available in these article pages. More detailed information can be found in the GS help pages [15].

When GS's bots are able to access the full text of the documents (either because the resource is openly available on the web, or thanks to the special agreements GS has with most commercial publishers), they also extract the list of cited references from each document. Thus, they are able to link citing and cited documents, which is how they can calculate citation counts.

When a list of cited references is parsed and processed, GS tries to find matches to those documents in its database. If it finds a match, it links the citing and cited document, and the cited document will have one more citation. However, if it doesn't find a match for a particular cited reference, the system will create a new bibliographic record of the type *[CITATION]*, to which the citing document will be linked. These records are also displayed in SERPs, although it is possible to exclude them, as their bibliographic information is often incomplete, and users won't be able to access their full-text. A *[CITATION]* record can become a full-fledged record if GS finds another version of the document on the web (because someone deposits it on a repository, or it becomes available from a publisher, etc.), and merges the two versions. Needless to say, this entire process is also completely automated.

Lastly, GS considers a wide range of parameters for ranking documents in the SERPs, such as "weighing the full text of each document, where it was published, who it was written by, as well as how often and how recently it has been cited in other scholarly literature" [16]. However, the detailed set of parameters and the weight each of them has in the ranking algorithm is not publicly available.

**2.2.2 What sources does Google Scholar index?**

The previous section makes clear the distinction between GS and traditional bibliographic databases. In their own words, "we index papers, not journals" [17]. However, this statement is only partially true. We'd rather define GS as a database that indexes web sources. Moreover, it deliberately includes some document collections (e.g., patents, court opinions, and *[CITATION]* type records).

GS crawls a wide variety of web sources, and indexes everything from those sources that it identifies as academic documents. That is why GS includes all documents regardless of subject, language, country or year of publication, and document type. The procedure GS follows to index new documents can be summarized in three steps, which are described below:

Step 1: Compilation of sources

Over the years, GS has compiled a huge list of sources, ranging from websites of academic institutions (higher education institutions, national research councils, commercial publishers, private companies, professional societies, non-governmental organizations, etc.), to other discovery tools (bibliographic databases, catalogues, directories, repositories, other search engines) available across the web. These sources, which are the most likely to host academic content, are the ones that shape the academic web. Once they add a source to their private master list, Google Scholar's spiders will visit it periodically to check whether new documents have been added, but also to verify that the documents indexed in the past are still available.

Besides the already mentioned academic sources, anyone can request that their website be considered for inclusion in GS. They are prepared to index websites that run in most of the common repository platforms (DSpace, EPrints…), journal platforms (OJS), and also simple personal websites.

Since GS's main objective is to facilitate content discovery, the sources must not require users to install additional applications, to log-in, use Flash, JavaScript, form-based navigation, or any other kind of unreasonable methods to access the documents. In addition to that, the website should not display popups, interstitial ads, or disclaimers. They specifically state that "all those websites that show log-in pages, error pages, or bare bibliographic data without abstracts will not be considered for inclusion and may be removed from GS".

Step 2: Document types

The next step is to index the academic documents available in each source. GS doesn't index all the documents in a source, only those that are academic in nature.

GS states that they cover mostly "scholarly articles, journal papers, conference papers, technical reports, or their drafts, dissertations, pre-prints, post-prints, or abstracts". However, content such as "news or magazine articles, book reviews, and editorials" is not considered appropriate for GS. However, appropriateness is not really a constraint, and the documentation also states that "shorter articles, such as book reviews, news sections, editorials, announcements and letters, may or may not be included".

These ambiguous declarations are a consequence of the automated way in which the system operates. Let's see a practical example:

The University of Oxford's official website (<ox.ac.uk>) can be considered a reliable source of academic information. GS has added this web domain to its master list of indexable sources (like it does with most universities). However, not all documents hosted in <ox.ac.uk> are of an academic nature. GS needs to automatically differentiate academic documents from all the rest. To do this, the system applies two approaches:

a) The parser approach: GS uses full-text parsers to identify the structure of documents. Taking advantage of the fact that many academic documents tend to present a fairly standardized structure (title, author names, abstract, body of the article, references…), detecting whether a document is academic or not is often possible, although errors do occur.
b) The location approach: GS automatically indexes all the documents hosted in specific locations where it is reasonable to expect that all documents will have an academic nature, i.e. institutional repositories.

For this reason, despite what is stated in Google Scholar's documentation, it is possible to find a great range of document types in GS. Documents are usually stored in the HTML or PDF format. Parsing the structure from these documents is not enough to detect their specific typology (article, book chapter, conference paper, etc.) when additional metadata is not available. Moreover, once it has been decided that all content from a given source will be indexed, the actual document type stops mattering. For example, we can find in GS many book reviews, a document type that is explicitly considered inappropriate (Figure 1).

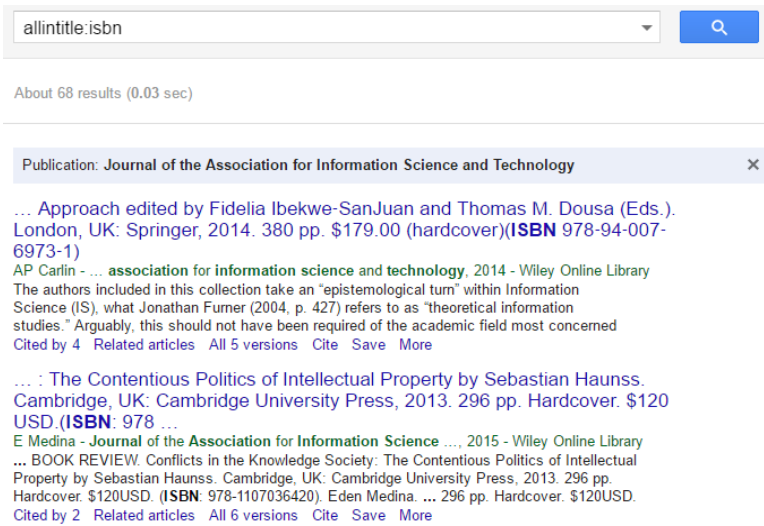

**Figure 1. Book reviews published in JASIST and indexed in Google Scholar**

Step 3: Documents

Lastly, documents themselves must also follow certain guidelines in order to be successfully indexed in GS. Some of them are compulsory, and others are only optional. Failure to comply with these rules may provoke an incorrect indexing of the documents, or, more likely, a complete exclusion from the search engine.

The system requires one URL per document (one intellectual work should not be divided into different files, while one URL should not contain various independent works). Additionally, the size of the files must not exceed 5MB. Although documents of a larger size can appear in GS, their full text (including cited references) will be excluded if they do not comply with this rule.

HTML and PDF are the recommended file types. Other document types such as DOC, PS, or PPT are also indexed, but they are a very small minority and they might not be processed as effectively as the others. Additionally, PDFs must follow two important rules. First, all PDF files must have searchable text. If the PDFs are just scanned images, the full texts (including the cited references) will not be processed, since Google Scholar's crawlers are unable to parse images. Second, all URLs pointing to PDF files must end with the .pdf file extension.

There are also rules regarding the description of the articles through metadata. Some fields are compulsory for all documents (title, authors, and publication date), while others are specific to each document type. Google Scholar supports HTML meta tags in various formats: Highwire Press, Eprints, BE Press, PRISM, and Dublin Core (the last one as a last resort, since there are no specific fields for journal title, volume, issue, and page numbers in this format). If no metadata is readily available in the HTML meta tags of the page describing the article, GS will try to extract bibliographic information by parsing the full text of the document directly. For this reason, GS also makes recommendations regarding the layout of the full texts:

- The title, authors, and abstract should all be in the first page of the text file.
- The title should be the first content in the document and no other text should be displayed with a larger font-size.

- The list of authors should be listed below the title, with smaller font-size, but larger than the font-size used for the normal text.
- At the end of the document, there should be a separate section called *References* or *Bibliography*, containing a list of numbered references.

Lastly, the abstract of the document should be visible to all users that visit the article page (regardless of whether they have access to the full text of the document or not) without needing to click on any additional links or log-in. If this requisite is not met, it is likely that the document will not be indexed in GS.

**2.2.3 Google Scholar's official bibliometric products**

GS has developed two secondary products that make use of both the bibliographic and citation data available in its core database. The first one (Google Scholar Citations) focuses on researchers, while the second one (Google Scholar Metrics) focuses on journals and articles. This section describes each product briefly and discusses the bibliometric indicators they provide.

**Google Scholar Citations**

Google Scholar Citations (GSC) was officially launched in November 16$^{th}$, 2011. This tool is an academic profile service meant to help researchers maintain an up-to-date list of their publications without much effort (it is updated automatically), and it also facilitates searches of people (rather than documents) who are experts in any given academic topic.

First, GSC profiles contain structured personal information (name of the researcher, affiliation, and research interests). Second, the profiles show a list of all the publications written by the researcher. For each of the publications, both bibliographic (authors, title, source, year of publication) and citation data (number of citations, and link to the list of citing documents) are offered. By default, documents are sorted decreasingly by number of citations, although they can also be sorted by year of publication or by title. Third, the profile also provides several author-level indicators (Table 1). These indicators are calculated considering two different timeframes: first, without any time restriction (useful for comparisons of senior scholars), and second, considering only citations received in the last five years (useful for comparisons of early-career researchers). It is important to keep in mind that these indicators are calculated automatically from the data in the publication list, without any sort of human supervision.

**Table 1. Google Scholar Citations' author-level metrics**

| Metrics | All | Last 5 years |
|---|---|---|
| **Citations** | Number of cites an author has received | Number of cites an author has received in the last 5 complete calendar years |
| **h-index** | The largest number h such that h publications have at least h citations | The largest number h such that h publications have at least h new citations in the last 5 years |
| **i10-index** | Number of publications with at least 10 citations | Number of publications with at least 10 citations in the last 5 years |

In GSC, users have access to a document search tool that enables them to find their publications by means of author name searches (it is possible to search as many name variants as necessary), or by known document searches (usually title searches). After all documents have been found, researches may merge versions of the same document that GS hasn't been able to detect, and fix bibliographic errors manually. All these operations only affect the researcher's profile, and not other co-authors' profiles, nor the results in GS Search.

The platform also offers additional services such as personalized alerts, lists of co-authors, areas of interests, list of authors by institution, etc. Therefore, authors can track the impact of their papers and other researchers' papers according to the data available in GS, and be instantly informed of new papers published by other authors. All these features make GSC a powerful and free research monitoring system.

This product may be viewed as a first step in the transition from an uncontrolled database to a better-structured system where authors, journals, institutions and areas of interest go through human filters. Nevertheless, the platform still lacks some essential features, like the identification of document types. This information can be defined by the owner of the profile in the document edit page, but it is not visible to other users who visit the profile. Another important issue is that author affiliations are not available at the level of documents. Although the affiliation field of the profile can be modified, it is only possible to display one affiliation at a time. However, researchers may change affiliations, and it wouldn't be fair to ascribe all documents by a researcher to only one institution if some of them were published while working at other institutions. Limitations like these diminish the usefulness of the platform for bibliometric studies (for more limitations please see section 1.4.2).

**Google Scholar Metrics**

Google Scholar Metrics (GSM) was launched in April 1$^{st}$ 2012 and can be defined as a hybrid between a bibliographic and bibliometric product that presents a ranking of journals according to bibliometric indicators calculated using citation data from recently published articles in those journals. If GS represented a paradigm shift in the market of bibliographic databases, GSM accomplished something similar with respect to journal rankings, especially when compared to products like Journal Citation Reports (JCR), Scimago Journal & Country Rank (SJR), or Journal Metrics [18]. GSM is an original product for various reasons:

*Inclusion policies*: GSM only covers journals which have published at least 100 articles in the last five years, and which have received at least one citation for any of those articles.

*Coverage*: Apart from journals, GSM also covers some conference proceedings from Computer Science and Electrical Engineering, and preprint repositories. Other typologies like court opinions, books, and dissertations are explicitly excluded.

*Sorting criteria*: Sources are sorted by ther h5-index (h-index for articles published in a given 5 year period). Using an h-index variant instead of a formula similar to the Journal Impact Factor (JCR), SJR, SNIP, or CiteScore (Journal Metrics) is probably one of the most distinct features of this product. A description of the indicators available in GSM is presented in Table 2. For each journal, only the articles which contribute to the h5-index are displayed (h5-core). These articles are also accompanied by their citation counts, and the list of citing documents to each article is also available.

**Table 2. Google Scholar Metrics**

| Metrics | All |
|---|---|
| **h5-index** | The largest number h such that at least h articles in that publication were cited at least h times each in the last five years |
| **h5-core** | The set of articles from a journal which a citation count above the h5-index threshold |
| **h5-median** | Median of the distribution of citations to the articles in the h5-core |

*Categorization of sources*: the first variable of categorization is the language of publication. In the version available at the time of this writing (launched in summer 2016, covering the period 2011-2015), the following languages were covered: English, Chinese, Portuguese, Spanish, German, Russian, French, Japanese, Korean, Polish, Ukrainian, and Indonesian. For each of these languages, except for English, the ranking displays the top 100 sources according to their h5-index. For English sources, a subject classification is also provided (Table 3). The classification scheme is made of 8 main categories and 302 subcategories. In each of the categories and subcategories, the number of results displayed is limited to the top 20 journals according to their h5-index.

**Table 3. Categories and number of subcategories in Google Scholar Metrics**

| Categories | Number Subcategories |
|---|---|
| **Business, Economics & Management** | 16 |
| **Chemical & Materials Science** | 18 |
| **Engineering & Computer Science** | 58 |
| **Health & Medical Science** | 69 |
| **Humanities, Literature & Arts** | 26 |
| **Life Sciences & Earth Sciences** | 39 |
| **Physics & Mathematics** | 24 |
| **Social Sciences** | 52 |

It should be pointed out that some subcategories are included in several categories (Library & Information Science, for example, is included both in Engineering & Computer Science, and in Social Sciences), and that one source may be included in more than one subcategory.

*Search tool*: The platform also provides an internal search box, which enables users to locate journals that are not included in any of the general rankings. Users can carry out keyword queries, which will return sources with names that match the query. Each response to a query contains a maximum of 20 sources, also sorted according to their h5-index.

The peculiar features of this journal ranking have been tested, finding numerous limitations, such as a lack of name standardization, irreproducible data, and a questionable mix of publication typologies [19]. Nevertheless, the product has improved since its first editions, revealing itself as a potential source for the evaluation of journals in the areas of Humanities and Social Sciences [20-21].

### 2.3 Radiographing a 'big data' bibliographic source

The goal of this section is to provide empirical data about the bibliographic properties of Google Scholar as a database. Three aspects will be discussed: size, coverage, and growth rate.

**2.3.1 Size**

One of the most crucial aspects that make us consider GS a *big data* source is the issue of determining its size. Unlike Scopus or WoS – highly controlled databases where finding out the total number of records only requires a simple query -, GS is a search engine that present what is available in the academic Web at a specific moment in time. However, the Web is not only dynamic, but also unstable and uncontrollable.

Therefore, the methodological difficulties of ascertaining the size of search engines are related to stability problems [22-23], precision of the results obtained [24-25], and the degree of permanence and persistence of the resources [26-28]. The high unstability of search results and the lack of precise search commands have led experts interested in finding the size of a search engine to use methods based on the extrapolation of frequencies of documents available in external sources [29].

As regards Google Scholar, there are two types of methods to calculate its size: direct methods (based on the execution of queries in the search engine), and indirect methods (estimations based on comparisons with external sources for which more information is known). Among the direct methods, three strategies are worth mentioning: web domain queries [30-32], year queries [3], and the so-called *absurd* queries [33]. Among the indirect methods, the capture/recapture method [34] and the proportion of documents in English respect to the total [33] have been attempted. For these last two methods, additional information must be known about the databases used as a reference. To date, these studies find that direct methods based on web domain queries and absurd queries are the ones that yield higher figures, similar in both cases.

Before discussing the calculation of the size of GS, it is appropriate to describe the characteristics of its coverage, since this is key aspect to understand the results. Google Scholar is currently made of two separate document collections: articles and case laws. The analysis of the latter is outside the scope of this chapter. The article collection is in turn divided into source documents, and cited references (documents that GS's crawlers have only been able to find as references inside other source documents or certain metadata-only bibliographic databases). Cited references are marked with the text *[CITATION]* in SERPs and can also accrue citations of their own, which are displayed in the same way as for source documents. There is a last document type to which GS gives special attention in its interface: patents.

The integration of source documents and cited references in the same list of results breaks from the way Web of Science and Scopus handle these types of records, where each collection is displayed separately. In WoS, cited references are accessible from a completely separate search system (Cited Reference Search), while in Scopus, cited references are also displayed separately as *secondary documents*.

There are two types of [CITATION] records: *linked citations* (documents for which only basic bibliographic information - but no access to full-text - has been found in some library catalogue or metadata-only database), and *unlinked citations* (documents that have been cited in source documents and which the system hasn't been able to find anywhere else on the Web).

Table 4 shows a compilation of the studies that have provided estimations of the size of Google Scholar. As can be expected, the results are affected by the estimation method, date of data collection, languages covered, and specific parameters of the searchers (inclusion or exclusion of cited references and patents).

**Table 4. Compilation of studies on the size of Google Scholar**

| AUTHORS | DATE | METHOD | COVERAGE | LANGUAGE | SIZE |
|---|---|---|---|---|---|
| **Aguillo (2012)** | August 2010 | Direct-Domains | Articles + citations+ patents | All | 86 million |
| **Ortega (2014)** | December 2012 | Direct- Date query | Articles + citations+ patents | All | 95 million |
| **Khabsa and Giles (2014)** | January 2013 | Indirect-Cap/Recap | n/a | English | 99 million |
| **Orduna-Malea et al (2015)** | May 2014 | Indirect-empirical studies | n/a | All | 171 million |
| | | Direct-Date query | Articles + citations+ patents | All | 100 million |
| | | Direct-Absurd query | Articles + citations+ patents | All | 170 million |
| **Aguillo (2017)** | January 2017 | Direct-Domains | Articles + citations+ patents | All | 194 million |
| **Delgado, Orduna-Malea, Martin-Martin (2017)** | March 2017 | Direct-Absurd query | Articles + citations+ patents | All | 331 million |
| | | | Articles | All | 184 million |
| | | Direct-Domains | Articles | All | 197 million |

Aiming to offer results as updated as posible, we replicated the absurd query in March of 2017. A series of year queries, combined with the command <-site:fsdfsdgsdh.info> were carried out, and the number of hits each search yielded was collected. Through this simple procedure, we obtained a total of 184,001,450 source documents. Together with cited references (134,160,570) and patents (13,742,920), bringing the total to 330,804,940 documents.

Figure 5 offers a comparison of the size of GS, WoS, and Scopus at two moments in time: 2011, and 2016. According to the most recent data, the coverage of Google Scholar seems to be almost three times as large as the coverage of WoScc (2.8:1) and Scopus (2.7:1).

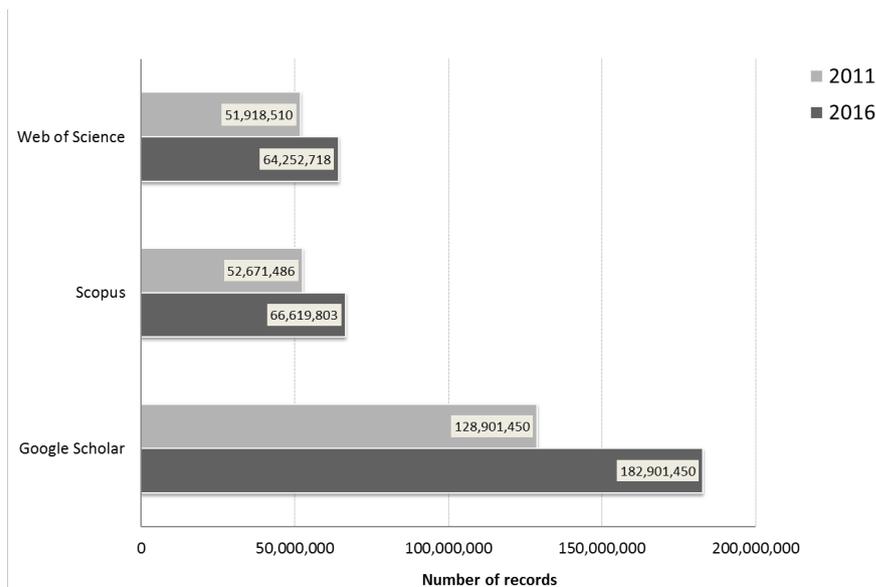

**Figure 2. Sectional coverage of Google Scholar, WoS Core Collection and Scopus (from origin to 2016, included)**

In order to test the robustness of the results, we compared the results returned by various types of queries to Google Scholar, carried out at different moments in time (Table 5). A high correlation was found among the results found for each year for all queries, regardless of whether cited references and patents were excluded or not.

**Table 5. Correlation among different queries (1800 – 2013)**

| Queries | Absurd Pure (2017) | Date Pure (2017) | Absurd Full (2017) | Absurd Full (2014) | Date Full (2014) |
|---|---|---|---|---|---|
| Absurd - Pure (2017) | 1 | 0.997 | 0.992 | 0.978 | 0.976 |
| Date - Pure (2017) | 0.997 | 1 | 0.994 | 0.984 | 0.983 |
| Absurd - Full (2017) | 0.992 | 0.994 | 1 | 0.990 | 0.990 |
| Absurd - Full (2014) | 0.978 | 0.984 | 0.990 | 1 | 0.995 |
| Date - Full (2014) | 0.976 | 0.983 | 0.990 | 0.995 | 1 |

Pure: excluding citations and patents; Full: including citations and patents

Obviously, all these results are merely approximations to the size of Google Scholar. Its exact size can't be ascertained with precision. Given the magnitude of the numbers the system returns (millions of documents), estimations are the best that we can expect when working with academic search engines.

Although the size of GS is clearly higher than that of other databases, the coverage by years can show us which database has a higher coverage in specific years. Figure 3 shows the number of documents by publication year that GS, WoS, and Scopus covered at the time of this writing.

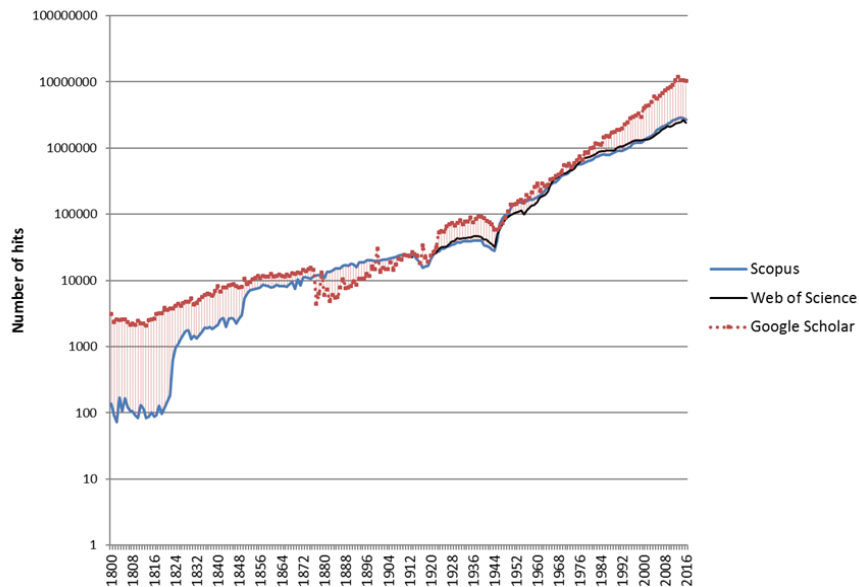

**Figure 3. Size of Google Scholar, WoS Core Collection and Scopus (1800 – 2016)**
Note: The size of Google Scholar is referred to citing sources, eliminating cited references, in order to make GS comparable with the remaining databases.

As Figure 3 shows, GS covers many more source documents than the other two databases, both for old material (first half of 19$^{th}$ century) and recent material (from the beginning of the 21$^{st}$ century onwards). The vast majority of the content covered by these databases has been published in the current century (70.4% of all documents).

Up to this point, we have analysed the size of GS by studying its source documents. However, it is also possible to approach this issue by studying the relationships between documents, that is, citations themselves. As was previously said, an indirect method to find out the size of GS is to use other databases as a benchmark. Differences in citation counts in documents that are covered both by GS and the database used as a reference can be considered an indication of the underlying differences of their document bases: if a database has been able to find more citations than the other database for a particular set of documents, its document base must be bigger (assuming the citation tracking mechanisms of the databases under comparison have roughly the same effectiveness).

Numerous studies have been published on this issue, most of them comparing GS to WoS and/or Scopus. The units of measurement used in these studies are varied: number of publications, number of citations, h indexes, and percentage of unique citations in each database. A simple summary of these empirical results can reveal to which degree the size of GS's document base is bigger than that of the traditional databases, and how these differences have changed over time.

A table with a list of 63 empirical studies that address the issue of the size of Google Scholar as compared to WoS and/or Scopus is available in the supplementary material [35] (see Appendixes I, III, and IV). For each study, the sample, discipline studied, unit of measurement used, results obtained, and ratio between the results for GS and the other databases. The results must be interpreted with caution, because the ratios depend to a large degree on the disciplines studied, the geographic scope (international or national), and the linguistic scope. Moreover, they can also be affected by the size of the samples (which in general tended to be very small).

From this meta-analysis, we can confirm several facts:

- To date, there are few studies that compare GS, WoS, and Scopus at the same time. The most frequent comparison is between GS and WoS.
- The vast majority of studies make comparisons on the basis of citations (54), way ahead of those that use documents – usually articles - (24), or indicators like the h index (12).
- Out of the 63 studies, only 8 yielded results where WoS Scopus surpassed GS in terms of size. All these studies analysed STEM fields, like Chemistry. The majority of studies show that, for any given set of documents covered both by GS and WoS/Scopus, the first is able to find a higher number of citations.
- The greatest differences in favor of GS are found in the Humanities and the Social Sciences. As regards STEM disciplines, GS is still able to find more citations, but the differences are less marked.
- The differences are greater when comparisons are made on the basis documents written in languages other than English.
- The differences between GS and the other databases seem to increase in the more recent studies, which might indicate an even broader coverage in GS respect to the other databases in recent years.
- Ratios of GS indicators to indicators from other databases are lower when comparisons are made on the basis of the h-index or the number of publications, and they are higher when citations are used instead. Ratios are even more in favor of GS when only unique citations are analysed (citations found in one database and not in the others). These unique citations in GS, that is, its ability to find citations that no other database is able to find (not limiting itself to strictly scientific sources, but covering all the academic and professional sources it can find), are what make GS truly unique.

So as to provide a broader and more updated perspective on this issue, we have analysed two large samples of documents covered both by GS and WoS. The results are consistent with the previous studies.

**Table 6. Compilation of self-elaborated materials on GS/WoS Citation Ratios at the document level**

| Source | Date of data collection | Description | N | GS citations | WoS citations | Ratio GS/WoS |
|---|---|---|---|---|---|---|
| **Self-elaborated. Publication forthcoming** | June-October 2016 | Articles and reviews with a DOI covered by WoS, published in 2009 or 2014. WoS data extracted from web interface | 2.32 million | 42.6 million | 27.6 million | 1.54 |
| **Self-elaborated for this chapter** | February 2017 | Highly cited documents in master sample. WoS data extracted only from GS/WoS integration | 69,261 | 80.8 million | 44.9 million | 1.80 |

According to the samples analysed in table 6, the ratio of GS citations to WoS citations ranges from 1.54 (for a sample of 2.32 million articles and reviews published in 2009 or 2014) to 1.80 (for a sample of 69,261 highly cited documents).

**2.3.2 Coverage**

After studying the size of Google Scholar, this section will focus on the characterisation of its content. To this end, its source, geographic, linguistic, discipline, and document type coverage will be discussed.

**Source coverage**

Unfortunately, neither Google Scholar nor its secondary products (GSM and GSC) provide a master list of sources. As previously said, GS is not a journal database, but a service that indexes academic documents from many web domains. For this reason, efforts to determine its sources should try to identify these web domains first.

The first exhaustive study of the sources covered by Google Scholar based on web domains was carried out by Aguillo [31], who concluded that the most frequent geographic top-level domain (ccTLD) was .cn (China), and that Harvard University was the Higher Education institution that contributed more content to Google Scholar. Ortega [3] went a step further by estimating the proportion of content provided by several types of content providers: publishers (41.6%), other Google products (22.5%), subject repositories (16.9%), and institutional repositories (11.8%). Martin-Martin et al [36] analyzed the sources of the primary versions of a set of 64,000 highly cited documents in GS, finding close to 6,000 content providers, among which the US National Institutes of Health (nih.gov), ResearchGate (researchgate.net), and Harvard University (harvard.edu) were the main providers of highly cited documents. This study also found that generic top-level domains (like .edu, .org, and .com) were more frequent that geographic TLDs. Lastly, Jamali y Nabavi [37], based on a series of topic queries, use Google Scholar to estimate the sources (researchgate.net, nih-gov) and top-level domains (.edu and .org) with a higher proportion of open access documents.

Aiming to offer more updated results, we carried out a series of *site:* queries in GS to find out the number hits returned for each of a list of 268 TLDs (251 geographic domains, and 17 first-generation generic domains). The searches were carried out in March 2017. Publication year restrictions weren't used, and cited references and patents were excluded.

Table 7 shows the main providers of documents according to theil top-level domain. China is first among the geographic TLDs (12.12% of the total content), and commercial companies (.com) lead the list of generic domains (45.39% of the total content).

**Table 7. Top 20 domain sources of Google Scholar (2017)**

| RANK | TLD | % | HCE | DESCRIPTION | TYPE |
|---|---|---|---|---|---|
| 1 | .com | 45.39 | 89,500,000 | Commercial | Generic |
| 2 | .org | 16.38 | 32,300,000 | Noncommercial | Generic |
| 3 | .cn | 12.12 | 23,900,000 | China | Country |
| 4 | .edu | 3.55 | 7,010,000 | US accredited postsecondary institutions | Generic |
| 5 | .jp | 3.40 | 6,700,000 | Japan | Country |
| 6 | .net | 2.06 | 4,070,000 | Network services | Generic |
| 7 | .ru | 1.72 | 3,400,000 | Russian Federation | Country |
| 8 | .gov | 1.69 | 3,340,000 | US Government | Generic |
| 9 | .br | 1.35 | 2,670,000 | Brazil | Country |
| 10 | .fr | 1.22 | 2,400,000 | France | Country |
| 11 | .kr | 0.94 | 1,850,000 | Korea Republic of | Country |
| 12 | .ua | 0.69 | 1,360,000 | Ukraine | Country |
| 13 | .id | 0.65 | 1,280,000 | Indonesia | Country |
| 14 | .es | 0.63 | 1,250,000 | Spain | Country |
| 15 | .pl | 0.56 | 1,110,000 | Poland | Country |
| 16 | .de | 0.51 | 1,010,000 | Germany | Country |
| 17 | .au | 0.44 | 864,000 | Australia | Country |
| 18 | .uk | 0.44 | 863,000 | United Kingdom | Country |
| 19 | .it | 0.40 | 797,000 | Italy | Country |
| 20 | .ca | 0.37 | 734,000 | Canada | Country |

Notes: cited references excluded; HCE: Hit Count Estimate

The sum of the results obtained for these 268 domains comes to 197,194,092 source documents, which is similar both to the one we obtained with the absurd query method in the previous section (184,001,450), and the one obtained by Aguillo [32], who used the same methodology (193,824,176). This reinforces the notion that the real number of source documents (excluding cited references) lies at around 200 million records. This figure is, however, a gross estimate, because there may be many duplicates, which will provoke an overestimation. At the same time, the fact that site command only counts primary versions would cause an infra-estimation if the web domain of the primary version does not match the web domain queried with the site search command.

Otherwise, the results indicate that most of the content is hosted in websites with generic (not geographic) top-level domains (69.7%), undoubtedly because of the weight of journal publishers, standalone journals, and American universities (.edu).

For the purpose of delving deeper into the issue of source typologies, we proceeded to calculate the size of five types of web domains. We wanted to illustrate the diversity and weight of the different types of sources from which GS feeds: digital libraries and bibliographic information systems (Table 8), publishers (Table 9), learned and professional societies (Table 10), US government agencies & international organizatios (Table 11), and universities (Table 12). These tables present the number of results found for each element, bot including and excluding cited references. The goal of these tables is to be able to observe which sources are generating a higher quantity of bibliographic records in GS. Lastly, it is worth remembering that these results only consider the primary versions of the documents (those GS has selected as primary versions), and therefore these tables shouldn't be understood as a ranking of sources sorted by size, but rather, a list that shows the diversity of sources available in GS.

Table 8. Digital Libraries (DL) and Bibliographic Information Systems (BIS) sources of Google Scholar

| | | | SOURCE | HITS | |
|---|---|---|---|---|---|
| RANK | DL & BIS | URL | TYPE | Citations excluded | Citations included |
| 1 | China National Knowledge Infrastructure | cnki.com.cn | Database | 17600000 | 19600000 |
| 2 | Google Books | books.google.com | Engine search | 3860000 | 9300000 |
| 3 | JSTOR | jstor.org | Digital library | 2920000 | 4680000 |
| 4 | Europe PubMed Central | europepmc.org | Subject repository | 2290000 | 4310000 |
| 5 | ResearchGate | researchgate.net | Social network | 2020000 | 2040000 |
| 6 | Proquest | proquest.com | Database | 1670000 | 1750000 |
| 7 | Astrophysics Data System | adsabs.harvard.edu | Database | 1510000 | 2040000 |
| 8 | J-STAGE | jstage.jst.go.jp | e-journal aggregator | 1460000 | 1750000 |
| 9 | Pubmed | ncbi.nlm.nih.gov | Subject Repository | 1360000 | 3350000 |
| 10 | Cyberleninka | cyberleninka.ru | Digital library | 1150000 | 1200000 |
| 11 | CAB Direct | cabdirect.org | Database | 1100000 | 1100000 |
| 12 | Refdoc | cat.inist.fr | Database | 1080000 | 2390000 |
| 13 | Academia.edu | academia.edu | Social network | 1020000 | 1030000 |
| 14 | CiteSeerX | citeseerx.ist.psu.edu | Search engine | 1010000 | 997000 |
| 15 | ERIC | eric.ed.gov | Database | 635000 | 695000 |
| 16 | AGRIS | agris.fao.org | Database | 537000 | 3620000 |
| 17 | Semantic Scholar | semanticscholar.org | Search engine | 526000 | 527000 |
| 18 | EBSCO | ebscohost.com | Database | 479000 | 479000 |
| 19 | Dialnet | dialnet.unirioja.es | Bibliographic portal | 458000 | 2280000 |
| 20 | ARXIV | arxiv.org | Subject repository | 403000 | 407000 |

**Table 9. Publisher sources of Google Scholar**

| RANK | PUBLISHERS | URL | HITS | |
|---|---|---|---|---|
| | | | Citations excluded | Citations Included |
| 1 | *Elsevier 2 | sciencedirect.com | 9340000 | 9410000 |
| 2 | John Wiley & Sons | wiley.com | 5960000 | 5970000 |
| 3 | Springer | springer.com | 5590000 | 5770000 |
| 4 | Taylor & Francis | tandfonline.com | 3200000 | 3240000 |
| 5 | Sage | sagepub.com | 1370000 | 1560000 |
| 6 | Lippincott Williams & Wilkins | lww.com | 1240000 | 1240000 |
| 7 | Cambridge University Press | cambridge.org | 1100000 | 1270000 |
| 8 | **Oxford University Press 2 | oxfordjournals.org | 951000 | 1240000 |
| 9 | Walter de Gruyter | degruyter.com | 595000 | 622000 |
| 10 | Nature Publishing Group | nature.com | 428000 | 458000 |
| 11 | Karger Publishers | karger.com | 347000 | 348000 |
| 12 | Chemical Abstracts Service | pubs.acs.org | 325000 | 325000 |
| 13 | BioMed Central | biomedcentral.com | 279000 | 279000 |
| 14 | Emerald | emeraldinsight.com | 236000 | 259000 |
| 15 | PLoS | journals.plos.org | 203000 | 203000 |
| 16 | World Scientific Publishing | worldscientific.com | 168000 | 170000 |
| 17 | Hindawi | hindawi.com | 167000 | 193000 |
| 18 | Elsevier 1 | elsevier.com | 108000 | 192000 |
| 19 | Inderscience Publishers | inderscienceonline.com | 82500 | 82500 |
| 20 | Brill | booksandjournals.brillonline.com | 68100 | 122000 |

* This Publisher owns another web domain (elsevier.com), in which we obtained 105,000 additional documents
** This Publisher owns another web domain (oup.com), in which we obtained 4,290 additional documents.

**Table 10. Learned & Professional societies of Google Scholar**

| RANK | Learned & professional societies | URL | HITS | |
|---|---|---|---|---|
| | | | Citations excluded | Citations included |
| 1 | Institute of Electrical and Electronics Engineers (IEEE) | ieee.org | 3410000 | 3650000 |

| 2 | Institute of Physics (IOP) | iop.org | 667000 | 702000 |
| 3 | Royal Society of Chemistry (RSC) | rsc.org | 470000 | 476000 |
| 4 | Association for Computing Machinery (ACM) | acm.org | 447000 | 601000 |
| 5 | American Phychological Association (APA) | apa.org | 406000 | 448000 |
| 6 | American Chemical Society (ACS) | acs.org | 326000 | 327000 |
| 7 | American Society of Microbiology | asm.org | 253000 | 256000 |
| 8 | International Union of Crystallography | iucr.org | 120000 | 122000 |
| 9 | American Mathematical Society (AMS) | ams.org | 93500 | 112000 |
| 10 | American Meterorological Society (AMS) | ametsoc.org | 59100 | 66300 |

**Table 11. Government agencies & International organizations sources of Google Scholar**

| RANK | Government agencies & International Organizations | URL | HITS Citations excluded | Citations included |
|---|---|---|---|---|
| 1 | National Institute of Informatics (NII) | nii.ac.jp | 2960000 | 13300000 |
| 2 | Japan Science & Technology Agency (JST) | jst.go.jp | 2740000 | 3060000 |
| 3 | US National Institute of Health (NIH) | nih.gov | 1420000 | 3380000 |
| 4 | Institut de l'information scientifique et technique | inist.fr | 1100000 | 2420000 |
| 5 | US Department of Education | ed.gov | 636000 | 696000 |
| 6 | Office of Scientific and Technical Information | osti.gov | 635000 | 1040000 |
| 7 | Food and Agricultural Organization (FAO) | fao.org | 545000 | 3510000 |
| 8 | Defense Technical Information Center | dtic.mil | 505000 | 644000 |
| 9 | National Aeronautics and Space Administration (NASA) | nasa.gov | 205000 | 252000 |
| 10 | National Criminal Justice Reference Service | ncjrs.gov | 120000 | 124000 |

**Table 12. University sources of Google Scholar**

| RANK | Universities | URL | HITS Citations excluded | Citations included |
|---|---|---|---|---|
| 1 | Harvard University | harvard.edu | 1410000 | 2170000 |
| 2 | Pennsylvania State University | psu.edu | 1030000 | 1080000 |
| 3 | Universidad de La Rioja | unirioja.es | 442000 | 2280000 |
| 4 | University of Chicago | uchicago.edu | 329000 | 346000 |
| 5 | Johns Hopkins University | jhu.edu | 324000 | 340000 |
| 6 | Universidade de São Paulo USP | usp.br | 155000 | 197000 |
| 7 | Masarykova Univerzita v Brně | muni.cz | 121000 | 125000 |
| 8 | Universiteit van Amsterdam | uva.nl | 105000 | 108000 |
| 9 | Universidad Complutense de Madrid | ucm.es | 105000 | 356000 |
| 10 | Helsingin yliopisto | helsinki.fi | 91400 | 142000 |

The results obtained in tables 8-12 illustrate the main sources from which Google Scholar feeds: big bibliographic information systems, including databases (Pubmed, Europe Pubmed Central, ADS), big commercial publishers (Elsevier, Springer y Wiley principalmente), other academic search engines (Semantic Scholar, Citeseer…), subject repositories (arXiv.org), social platforms (ResearchGate, Academia.edu), as well as Google's own book platform (Google Books). Additionally, research government agencies (like the Japanese National Institute of Informatics, and the Japan Science and Technology Agency), professional associations (IEEE), and universities (Harvard University is still leads this group). It should be noted that the results of some institutions (especially universities) are influenced by the existence of bibliographic products that are hosted within these universities' domains (Dialnet in La Universidad de La Rioja, CiteseerX in Pennsylvania State University, AGRIS in FAO, ERIC in the US Department of Education, etc.).

However, these results should be interpreted with caution, because the methodology used to collect the data has several limitations: all hit counts displayed by Google Scholar (and Google, for that matter) are only approximations, not exact figures. What's more, the *site:* command is not exhaustive either: it only works with the primary versions of the documents in GS. Google Scholar implements a procedure to group together all the versions of a same document that may be available on the Web [38]: subject and institutional repositories, the author's personal website, a social platform, and the official *version of record* available in a journal or publisher's website. From all the versions found by the search engine, one of them (usually the version of record, if there is one) is selected as the primary version. The rest of the versions can be found under the *All x versions* link available below each record.

This means that, when a search containing the *site:* command is carried out, the system will only return the records in which the source of the primary version matches the searched source, even though there may be many more records from that source that haven't been considered primary versions. If we focus on the case of ResearchGate, we can see that the 2,020,000 documents found (Table 8) are very far from the over 100 million documents the company claimed to cover in March 2017 [39]. This difference can be explained in part by the documents that are covered as secondary and not primary versions. For these reasons, the results in tables 8-12 are most likely an underestimation of its real coverage, although it provides important clues as to the main sources it indexes.

**Geographic coverage**

The geographic coverage of the documents covered by GS is also difficult to analyse, because the system is not designed to carry out searches based on authors' institutional affiliations. Therefore, like with the source coverage, a possible although biased approach is to analyse the distribution of geographic domains. Orduna-Malea and Delgado Lopez-Cozar, using this methodology [40], found that the domains for the United States, Chinar, and Japan were the ones that yielded the highest hit counts estimates (HCE), which is consistent with the results obtained by Aguillo [31]. Table 13 shows the top 10 geographic domains according to their HCE, and their evolution in the last 6 years [31; 40].

**Table 13. Top 10 geographic domains of Google Scholar**

| Country | TLD | Hit Counts Estimate | | | % |
|---|---|---|---|---|---|
| | | 2010 | 2013 | 2016 | |
| **China** | .cn | 7,520,000 | 30,700,000 | 23,900,000 | 12.12 |
| **USA** | * | N/A | 16,019,000 | 10,943,500 | 5.55 |
| **Japan** | .jp | 1,720,000 | 10,400,000 | 6,700,000 | 3.40 |
| **Russia** | .ru | 995,000 | N/A | 3,400,000 | 1.72 |
| **Brazil** | .br | 1,440,000 | 2,320,000 | 2,670,000 | 1.35 |
| **France** | .fr | 2,820,000 | 4,210,000 | 2,400,000 | 1.22 |
| **South Korea** | .kr | 481,000 | 1,720,000 | 1,850,000 | 0.94 |
| **Ukraine** | .ua | 210,000 | N/A | 1,360,000 | 0.69 |
| **Indonesia** | .id | N/A | N/A | 1,280,000 | 0.65 |
| **Spain** | .es | 907,000 | 2,990,000 | 1,250,000 | 0.63 |
| **Poland** | .pl | 220,000 | N/A | 1,110,000 | 0.56 |

2010 data: from [31]; 2013 data: from [40]; 2016 data: self-elaborated for this chapter. In all cases citations are excluded; N/A: Not Available.
* US data is obtained by merging the results from .us, .mil, .edu and .gov webdomains.

This strategy, however, is imprecise, because it doesn't consider generic top-level domains like .com and .org, which are precisely the ones that are most used. This explains why the United States are clearly underrepresented, since a great proportion of .com domains belong to institutions from this country [41].

There are few studies on the geographic distribution of documents in GS, and those few that have been published focus on the geographical origin of the journals indexed in Google Scholar Metrics, and the comparison of these journals with the ones covered by WoS and Scopus in specific disciplines like Communication [42], Nursing [43], and Library and Information Science [44]. GS and GSM seem to get closer to the actual distribution of scientific journals by country of publication than the other databases.

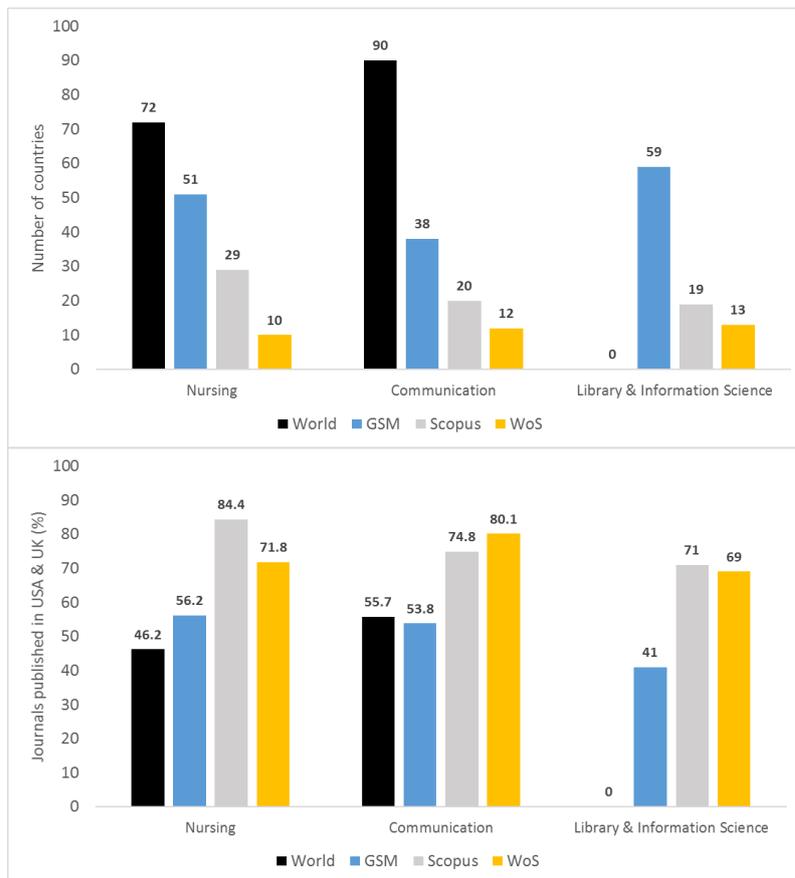

**Figure 4. Nursing, Communication, and Library & Information Science journals published according to Google Scholar Metrics, Scopus, and WoS data (up) Number of countries where the journals are published (bottom) Percentage of journals that are published in the USA or UK**
Note: In the case of Library & Information Science journals, Google Scholar is used instead of Google Scholar Metrics; and no data is available for World category.

Similar results are found in a study of over 9,000 Arts, Humanities, and Social Sciences journals indexed in GSM, where GSM not only covers journals from more countries, but the English-language bias is clearly less pronounced than in Scopus and WoS (Figure 5).

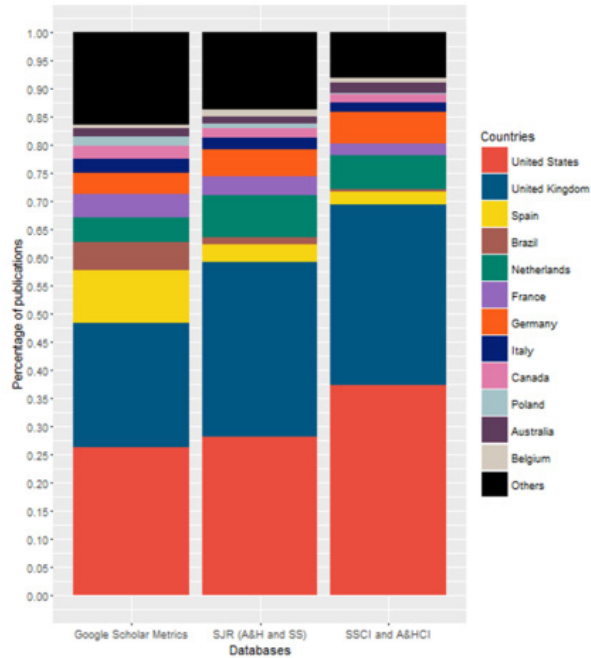

**Figure 5. Distribution of countries of publication in Arts, Humanities, and Social Sciences journals covered by GSM, SJR (Scopus), and JCR (Web of Science) Linguistic coverage**

It is also difficult to find out the language distribution of the documents covered by GS, because the options the interface offers to filter by language are very limited. Users can limit search results to one or several of the following languages: Chinese (simplified and traditional), Dutch, English, French, German, Italian, Japanese, Korean, Polish, Portuguese, Spanish, and Turkish. However, users have to navigate to the settings page to find these options, they are not available from the main search interface.

In a study carried out by Orduna-Malea and Delgado [40], which analysed the quantity of records by language in WoS, Scopus, and Google Scholar, a very high percentage of documents written in English was found in the first two databases (90%), while GS offered a higher linguistic diversity because it covered other languages (especially Italian, Spanish, French, and Japanese). A year later, Orduna-Malea et al. [33], based on an analysis of the empirical studies on this topic, estimated that documents in the English language represented approximately 65% of all documents in GS.

Studies on the publication languages of journals available in GSM, as compared to those available in WoS or Scopus, reach the same conclusions. Both in Communication journals [42], Nursing journals [43], and Library and Information Science journals [44], GSM is closer to the actual distribution of languages used in scientific journals around the world, thus overcoming the bias towards English-language sources that prevails in Scopus and WoS (Table 14). While in the latter two the proportion of English-language sources ranges from 80 to 93% of all sources, in GSM this figure is much lower: between 61% and 65% (Table 15).

**Table 14. Number of different languages in Nursing and Communication journals indexed in Google Scholar Metrics, Scopus, and WoS**

| Discipline | World | GSM | Scopus | WoS |
|---|---|---|---|---|
| **Nursing** | 33 | 20 | 13 | 6 |
| **Communication** | 23 | 14 | 7 | 6 |

**Table 15. Percentage of journals published in English in Nursing and Communication, and indexed in Google Scholar Metrics, Scopus, and WoS**

| Discipline | World | GSM | Scopus | WoS |
|---|---|---|---|---|
| **Nursing** | 57.2 | 61.9 | 81.2 | 92.7 |
| **Communication** | 70 | 65.3 | 91.6 | 87.8 |

An analysis of the Arts, Humanities, and Social Sciences journals available in the 2010-2014 edition of GSM yields similar results: GSM no only covers journals written in more languages but its English-language bias is also lower than in the other two databases (Figure 6), in spite of the fact that the study focused only on journals with titles written in latin characters. GSM, because of its inclusion criteria, doesn't cover many journals for which articles are available in GS.

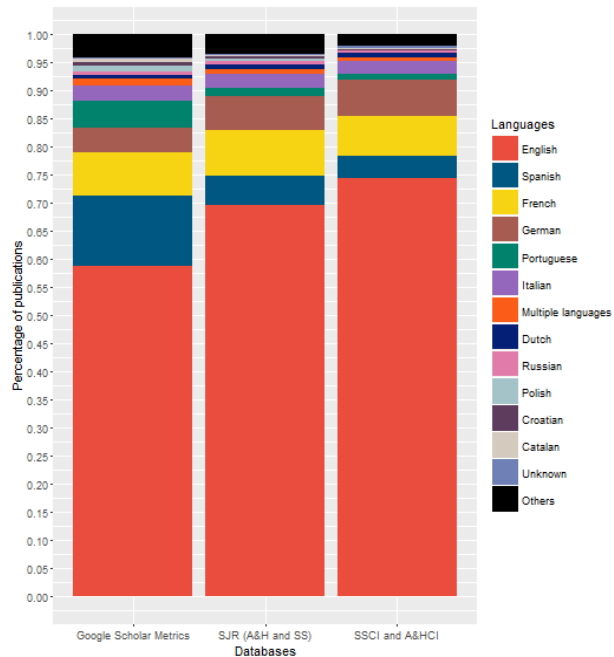

**Figure 6. Distribution of Arts, Humanities and Social Sciences journals indexed in GSM (2010-2014) as compared to SCImago journal Rank (AH&SS only), and Journal Citation Reports (SSCI & A&HCI)**

Aiming to obtain additional empirical data about the linguistic distribution of the content available in Google Scholar, we carried out a series of searches in GS. For each of the 12 languages that GS allows users to choose from to limit the search results (Simplified Chinese, Traditional Chinese, Dutch, English, French, German, Italian, Japanese, Korean, Polish, Portuguese, Spanish, and Turkish), 67 keyword-free, publication year queries were carried out, one for each publication year for the period 1950-2016 (871 queries in total).

Table 16 shows the distribution of results by language (excluding cited references and patents). As can be seen, English-language results make up half of the total amount of results (49.8%), followed by the sum of simpled and traditional Chinese results (33.7%). These results are consistent with the figures on geographic coverage through the analysis of web domains presented previously, and confirm the preeminence of the United States and China in Google Scholar's coverage.

**Table 16. Distribution of languages in Google Scholar results (cited references and patents excluded)**

| LANGUAGE | DOCUMENTS | % |
|---|---|---|
| English | 90,932,140 | 49.76 |
| Chinese | 61,545,203 | 33.70 |
| Japanese | 6,327,073 | 3.46 |
| German | 4,326,244 | 2.37 |
| Spanish | 4,144,354 | 2.27 |
| French | 3,657,705 | 2.00 |
| Portuguese | 2,403,898 | 1.32 |
| Korean | 2,131,744 | 1.17 |
| Italian | 999,134 | 0.55 |
| Polish | 766,266 | 0.42 |
| Dutch | 475,703 | 0.26 |
| Turkish | 472,830 | 0.26 |
| Other | 4,534,156 | 2.48 |
| **TOTAL** | **182,716,450** | **100** |

Even considering the disproportionately huge amount of English and Chinese results (clearly influenced by the sources from which Google Scholar extracts data), the distribution of the other languages is unquestionably more varied than the distribution presented by other databases. Figure 7 shows the relative distribution of languages in Google Scholar, Scopus, and WoS.

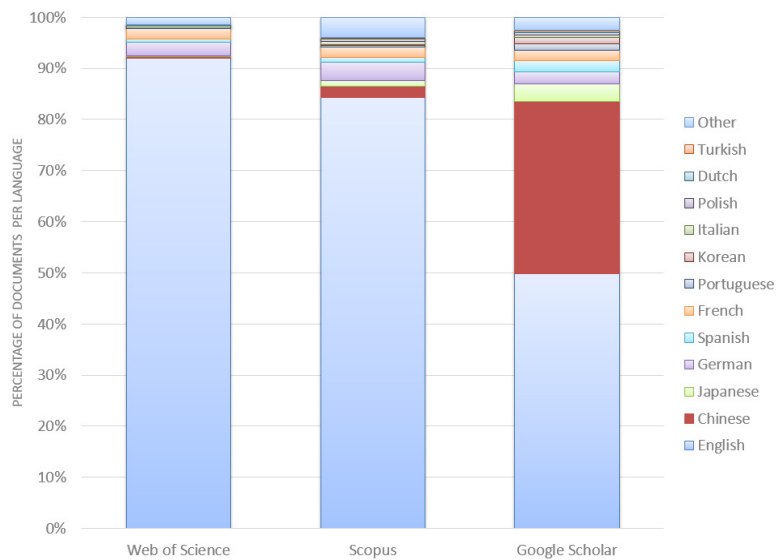

**Figure 7. Distribution of the languages of documents indexed in Google Scholar, Scopus, and WoS (1800-2016)**

While the percentage of documents published in English in WoS and Scopus is of 90% and 80% respectively, in Google Scholar the percentage is closer to 50%, and therefore the rest of languages are noticeably better represented.

Again, we'd like to warn that the results obtained must be interpreted in the context of a search engine, and not a bibliographic database. Google Scholar automatically identifies the language of a document from certain parameters. However, a document might contain text in several languages. Therefore, the same document might be classified in various languages. Additionally, in some cases the fact that a document is hosted in a geographic domain can help Google Scholar identify its language (for example, .cn is associated with the Chinese language), even if sometimes the documents are not written in the expected language. For those reasons, the number of results might be an overestimation.

**Discipline coverage**

Discipline coverage is another crucial aspect of the analysis of a bibliographic database. However, studies published to date mostly deal with journal coverage in GS as regards specific disciplines and countries.

From the data available about Spanish journals in the areas of Social Sciences, covered by GS and GSM IN 2011 [45], and the data available in IN-RECS [46], a ranking of Spanish journals in disciplines related to the Social Sciences, a very informative table about the coverage of GS and GSM was developed (Table 17). Of the 1,090 Spanish journals studied in IN-RECS, 95.2% (1,038) were covered by GS. What's more, for some disciplines, GS covered even more journals than IN-RECS. If we extrapolate those results, we can estimate that GS is very close to covering all active scientific journals. Of course, this hypothesis wouls require a more varied sample of journals to be tested.

On the other hand, GSM covers just over a third of all active Spanish Social Sciences journals (36.97%). This is most likely caused by the inclusion criteria enforced by the system (journals must have published at least 100 articles in the last 5 years, and received at least one citation). Nevertheless, as previously shown (3.2.2, and 3.2.3), even with those limitations, GSM still covers many more journals than traditional databases (WoS and Scopus), and, therefore, is able to display a much broader spectrum of disciplines, better representing the scientific landscape.

**Table 17. Coverage of Spanish journals by discipline in IN-RECS, Google Scholar, and Google Scholar Metrics**

| Discipline | IN-RECS | Google Scholar | | Google Scholar Metrics | |
|---|---|---|---|---|---|
| | Year (2010) | Year (2011) | Coverage (%) | Year (2011) | Coverage (%) |
| Law | 341 | 251 | 74 | 110 | 43.8 |
| Education | 166 | 157 | 95 | 69 | 43.9 |
| Economy | 136 | 137 | 101 | 55 | 40.1 |
| Psychology | 108 | 109 | 101 | 42 | 38.5 |
| Sociology | 82 | 87 | 106 | 25 | 28.7 |
| Political science and Administration | 60 | 56 | 93 | 21 | 37.5 |
| Geography | 51 | 54 | 106 | 15 | 27.8 |
| Anthropology | 46 | 46 | 100 | 10 | 21.7 |
| Sport | N/A | 42 | N/A | 14 | 33.3 |
| Urban studies | 43 | 39 | 91 | 15 | 38.5 |
| Library and Information Science | 33 | 36 | 109 | 15 | 41.7 |
| Communication | 24 | 24 | 100 | 12 | 50.0 |
| **TOTAL** | **1090** | **1038** | **95** | **403** | **37.0** |

In a study focused on the quantity of Spanish journals indexed in GSM, as compared with the total number of active Spanish journals, which according to the Ulrichs directory is around 2,668 journals [47], similar results are obtained. Only 48.7% (1,299) of Spanish journals were found in GSM (Figure 8), but if this result is compared to the traditional journal rankings, we find that, in spite of its inclusion criteria, GSM covers twice the amount of journals that SJR (Scopus data), and ten times more than the Journal Citations Reports (WoS data). However, there are studies that yield different results. Gu and Blackmore [48] find that, of a sample of 41,787 refereed academic journals from all disciplines covered by the Ulrichs directory, only 20.8% (10,354) were to be found in GSM, a lower amount than those that were found in SJR (32%, 15,911).

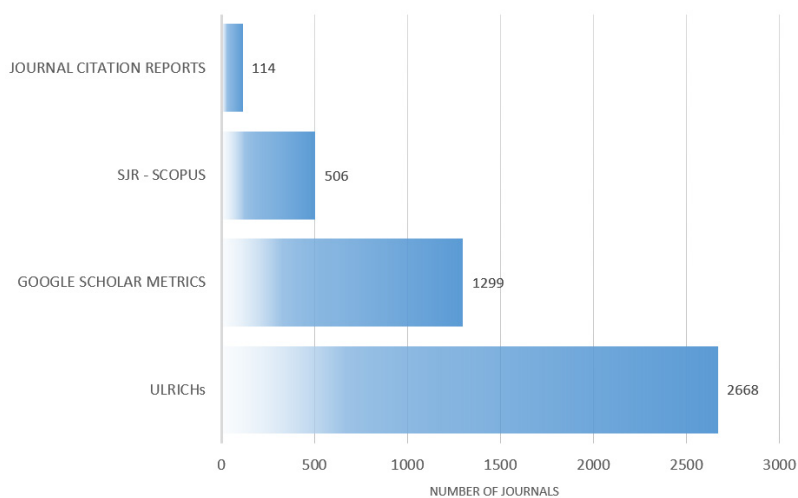

**Figure 8. Number of Spanish scientific journals covered by Journal Citation Reports, Scopus, Google Scholar Metrics, and Ulrichs directory.**

An additional method to quantify the number of publications covered by databases is to analyze the number of citations received by a sample of documents in a specific scientific discipline. With this approach we can find studies focused on journals [49-50] and researchers [51-54]. There is one other approach, based on the characterization of the documents returned by GS to sets of topic queries [37].

Important disciplinary differences can be observed in the list of studies that offer empirical evidences on the functioning of Google Scholar. While in the Social Sciences, the Humanities, and Engineering (especially in Computer Science) the ratio of citations in GS to citations in other sources is very high, the differences are much lower when STEM fields are analyzed. As was previously commented, the field of Chemistry was initially covered poorly in GS, mostly due to the refusal of the American Chemical Society and other important publishers to be indexed by the search engine. These problems have already been solved [55; 51].

**Document type coverage**

Lastly, the last aspect of the coverage of GS has to do with document typologies. In order to learn about the wealth and diversity of document sources from which GS extracts data, we must summarise the results offered by the studies that have analysed the distribution of citing documents according to their typology in several databases (Figure 9). The main feature of Google Scholar is that it indexes more diverse document typologies. Indeed, it is the database where journal articles constitute a lower percentage of the total documents (from 28% to 70% depending on the samples). Conversely, in WoS and Scopus journal articles make up 90% of the documents, which means there is a very limited coverage of conference communications and books.

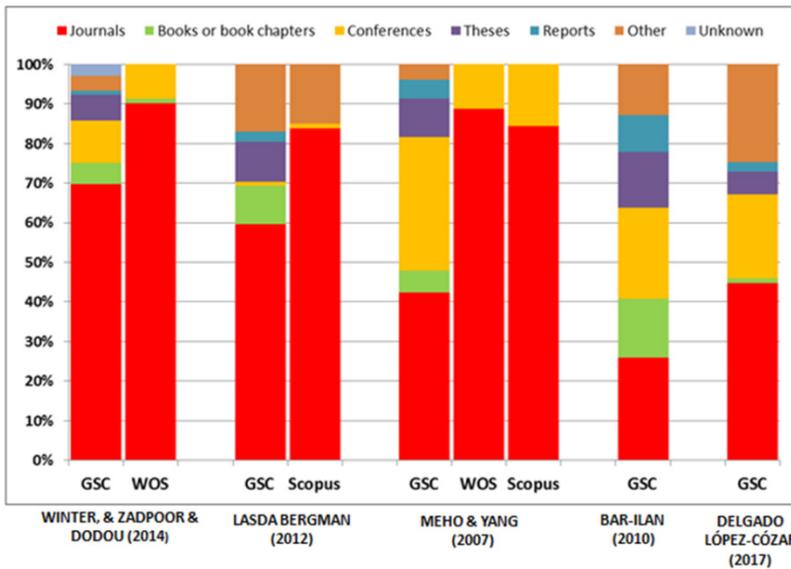

**Figure 9. Comparison of the document type distributions in GS, WoS, and Scopus as found in several studies**

Aiming to obtain more updated and representative results, we proceeded to carry out a similar series of queries to those carried out for the analysis in section 3.2.3. In this case, however, cited references and patents were included in the search. 871 queries (by language and year of publication) were carried out, extracting the 1,000 search results available for each query. 861,843 documents were extracted [56].

Given that GS doesn't provide information about the typology of the documents that are displayed in SERPs, document type identification becomes a rather complex task. By means of matching techniques to other data sources (WoS, CrossRef), and by applying a set of heuristics to extract more metadata from the websites that hosted the documents, the typology of 53.8% of the documents in the sample was ascertained (398,549). Table 18 shows the number of documents by typology and language.

**Table 18. Distribution of document typologies in Google Scholar (N=861,839 documents)**

| Document type | TOT | EN | CN | JP | DE | ES | FR | PT | KO | IT | PL | NL | TR |
|---|---|---|---|---|---|---|---|---|---|---|---|---|---|
| Unknown | 463290 | 11299 | 85547 | 29313 | 32092 | 40926 | 27375 | 38655 | 26206 | 38773 | 41803 | 44671 | 46630 |
| Article | 260211 | 39264 | 38550 | 34416 | 16087 | 10384 | 21357 | 15998 | 33606 | 7584 | 13638 | 13499 | 15828 |
| Book / Chapter | 120304 | 14418 | 5637 | 1912 | 18461 | 14035 | 15593 | 11559 | 311 | 19702 | 11054 | 6233 | 1389 |
| Thesis | 10919 | 33 | 3912 | 590 | 117 | 74 | 347 | 244 | 2891 | 35 | 35 | 702 | 1939 |
| Conference | 5741 | 718 | 162 | 447 | 159 | 375 | 626 | 505 | 240 | 587 | 468 | 439 | 1015 |
| Other | 649 | 20 | 112 | 305 | 2 | 3 | 62 | 2 | 93 | 2 | 0 | 1 | 47 |
| Report | 381 | 29 | 78 | 14 | 25 | 13 | 36 | 8 | 0 | 10 | 0 | 166 | 2 |
| Unpublished | 183 | 12 | 2 | 0 | 60 | 6 | 4 | 29 | 0 | 31 | 0 | 29 | 10 |
| Patent | 161 | 136 | 0 | 3 | 0 | 20 | 0 | 0 | 0 | 2 | 0 | 0 | 0 |
| **TOTAL** | 861839 | 65929 | 134000 | 67000 | 67003 | 65836 | 65400 | 67000 | 63347 | 66726 | 66998 | 65740 | 66860 |

EN: English; CN: Chinese; JP: Japanese; DE: German; ES: Spanish; FR: French; PT: Portuguese; KO: Korean; IT: Italian; PL: Polish; NL: Dutch; TR: Turkish

If we focus only on the documents for which the typology could be identified, articles (65.3%) and books (30.2%) are the most frequent typologies, followed at a distance by doctoral theses and conference communications. However, we must bear in mind the limitations of the sample. The search strategy used (keyword-free searches which only limited by year and language of publication), combined with Google Scholar's limitations (a maximum of 1,000 results per query, sorted by GS's relevance ranking algorithm), result in a sample biased towards highly cited documents, because the number of citations is the parameter that weighs the most in these type of queries [57].

In order to illustrate the bibliographic diversity in GS, Table 19 provides a list of the document types analysed in the empirical studies that have addressed this issue.

**Table 19. Document typologies found in Google Scholar**

| Bachelor's Dissertations | Bibliography | Biographical item | Blog |
|---|---|---|---|
| Book reviews | Books | Book chapters | Book reviews |
| Conference paper proposals | Conference papers | Conference posters | Conference presentations |
| Conference keynotes | Doctoral dissertation proposals | Doctoral dissertations | Doctoral qualifying examinations |
| Editorial | Guidelines and Clinical Algorithm | Interviews | Journal articles |
| Letters to the editor | Master's thesis proposals | Master's Theses | Civil service competitive examination reports |
| Notes | Preprints | Presentation Slides | Regulations |
| Reports | Research proposals | Research reports | Reviews |
| Series | Short Survey | Student portfolios | Supplementary Material |
| Syllabi | Term papers | Tweets | Unpublished Manuscripts |
| Unpublished papers | Web Pages | Web documents | Working papers |
| Workshop papers | Yearbooks | | |

Clearly, Google Scholar's bibliographic wealth is due to the manner in which it indexes information: the search engine indexes any document that is hosted in the academic web, providing it meets certain technical and structural criteria. The consequence, at any rate, is that the presence of full text conference proceedings, book chapters, reports, patents, presentation slides (either from university courses, conferences, or other events), and especially monographs and doctoral theses make of Google Scholar a unique tool not only to find information, but also to find citation data that is not available anywhere else.

**2.3.3 Growth rate**

Although sectin 1.3.1 already presented some results as to this search engine's growth rate (even comparing it to Scopus and WoS), a sectional approach such as the one represented in Figure 2 misses the main properties of GS, such as its dynamic nature. All content (both source documents and their citations, new or old) in GS is updated automatically.

As regards longitudinal analyses, Harzing [51; 58] studies the growth of citations to 20 Nober Prize winners in four disciplines, detecting a growth of 4.6% from April 2012 to April 2013. A similar result (4.4%) is found by Harzing y Alakangas [52], where 146 senior researchers from the University of Melbourne were analysed.

Retroactive growth (that is, the inclusion of documents published a long time ago) was in part addressed by De Winter et al [50], who analysed the relative difference between citation counts to a classic article up to mid-2005 measured in mid-2005 and citation counts up to mid-2005 measured in April 2013.

The speed with which Google Scholar indexes new source documents (and finds more citations to documents already in its document base), was addressed by Moed, Bar-Ilan, and Halevi [54]. The authors compute the indexing speed of Google Scholar for 12 journals in 6 different disciplines, and compare the results to those found in Scopus. Although there are differences among disciplines and results are affected by the Open Access policies of big publishers, the authors find that "the median difference in delay between GS and Scopus of indexing documents in Scopus-covered journals is about 2 months". The latest study on this issue to date was written by Thelwall and Kousha [59], which focuses on early citations to journal articles. The selected a sample of articles published in LIS journals between January 2016 and March 2017. The results in Google Scholar are compared to those in WoS, Scopus, and ResearchGate. The results in this study show that GS clearly outperforms all the other databases in terms of finding early citations, although ResearchGate's data is quickly becoming an interesting source for citation data as well.

What follows is a small-scale analysis that aims to illustrate this phenomenon, which has consequences not only for document searches, but in the speed with which citations are detected in the system. Articles accepted by the Journal of the Association for Information Science and Technology (JASIST), and made available as advance online publications between January 1$^{st}$ and March 25$^{th}$ 2017 were identified, noting the specific date when they were made available online. Secondly, those articles were searched in GS, and, in the cases when they were found, the exact date of indexing was saved (this information is available when documents are sorted by date). Knowing these two dates, we were able to compute the speed of indexing (number of days since the article was first available online, until GS picked it up). Results are displayed in Table 20.

**Table 20. Speed of indexing for JASIST articles in Google Scholar**

| Article | Online Publication | Scopus Index | Wos Index | Google Scholar | | | Online Age | Index Speed |
| --- | --- | --- | --- | --- | --- | --- | --- | --- |
| | | | | Index | Other Version | Days since index | | |
| 1 | 20-ene | YES | NO | NO | YES | | | |
| 2 | 24-ene | YES | NO | NO | YES | | | |
| 3 | 27-ene | YES | NO | YES | NO | 56 | 58 | 2 |
| 4 | 21-feb | *NO | NO | YES | NO | 31 | 33 | 2 |
| 5 | 21-feb | YES | NO | YES | NO | 31 | 33 | 2 |
| 6 | 27-feb | YES | NO | NO | YES | | 27 | |
| 7 | 27-feb | YES | NO | YES | NO | 26 | 27 | 1 |
| 8 | 27-feb | *NO | NO | YES | NO | 26 | 27 | 1 |
| 9 | 27-feb | YES | NO | YES | NO | 26 | 27 | 1 |
| 10 | 27-feb | YES | NO | YES | YES | n/a | 27 | |
| 11 | 07-mar | YES | NO | YES | NO | 17 | 19 | 2 |
| 12 | 07-mar | YES | NO | NO | YES | | 19 | |
| 13 | 13-mar | YES | NO | NO | YES | | 13 | |
| 14 | 20-mar | NO | NO | YES | NO | n/a | 6 | |
| 15 | 20-mar | NO | NO | NO | YES | | 6 | |
| 16 | 20-mar | NO | NO | YES | NO | 3 | 6 | 3 |
| 17 | 20-mar | NO | NO | YES | NO | 3 | 6 | 3 |
| 18 | 20-mar | NO | NO | YES | YES | n/a | 6 | |
| 19 | 20-mar | NO | NO | YES | NO | 3 | 6 | 3 |
| 20 | 20-mar | NO | NO | NO | YES | | 6 | |
| 21 | 20-mar | NO | NO | NO | YES | | 6 | |
| 22 | 25-mar | NO | NO | NO | YES | n/a | 1 | |
| 23 | 25-mar | NO | NO | NO | NO | | 1 | |
| 24 | 25-mar | NO | NO | NO | NO | | 1 | |

\* Not indexed in Scopus, because they are book reviews

On the date of data collection (March 26th) GS had indexed 13 out of 24 of the articles analysed from the publisher's website, although in 4 cases the date of indexing in GS was not available, because these documents were previously available in GS as preprints. What's more, out of the 11 articles GS had still not picked up from the publisher's website, 9 were available from other sources (mainly subject or institutional repositories). Only the two most recent articles (available in the publisher's website only one day before the analysis was carried out) weren't available in GS in any form. As regards the indexing speed, it ranges from one to three days. It is worth taking into account that there is a $\pm 2$ day margin of error, because we know the date of indexing but not the exact hour. According to these data, it seems that it only takes around two days for documents published in JASIST to be indexed in Google Scholar, although a larger sample would need to be analysed to confirm this for sure.

If we compare these results to the coverage of these documents in other databases, we can observe that Scopus had indexed the documents that had been made available up to March 13th. Although the exact date of indexing in Scopus is unknown, it was necessarily below 13, a very respectable speed considering it is a controlled database. However, these documents are classified as *in press* in the platform, and Scopus doesn't compute citations to documents until they are formally published in a journal issue, something that can take months. WoS doesn't index documents until they formally published, and therefore doesn't cover any of the documents in the sample.

**2.4 Google Scholar's data for scientometric analyses**

Lastly, this chapter wouldn't be complete without addressing the limitations of this search engine as a source of data for bibliometric analyses.

It is important to differentiate between the limitations from which this platform suffers by design for a specific purpose, and the various kinds of errors that the search engine makes when it processes data from the academic web. Errors are deviations from the expected or normal functioning of the tool (like for example the existence of duplicate citations, versions of the same document that haven't been merged, incorrect or incomplete attribution of authorship, etc.). Limitations, on the other hand, refer to the characteristics that can compromise the suitability of the tool for a specific purpose, especially if it is not the original purpose for which the tool was first developed (like for example, using Google Scholar as a source of data for bibliometric analyses, instead of as a search tool).

**2.4.1 Errors in Google Scholar**

The studies that have been published on the topic of errors found in GS are rather disorganized and superficial. There are few empirical evidences, and they often lack proper systematic study backed by representative samples. Most of the time, only anecdotal evidence is presented, without addressing the important issue of the degree of pervasiveness of the errors (how often the errors occur throughout the document base). The results in these studies are difficult to summarise and compare, and they become obsolete very quickly, because GS is being constantly updated and introduces improvements to its algorithms regularly.

Following on the footsteps of the numerous and sharp studies carried out by Jacsó [12; 30; 60-68], below we present a taxonomy of the types of errors made by GS. We propose to divide errors in four broad groups (search-related errors, parsing-related errors, matching errors, and source-related errors), described in Table 21. These errors can affect bibliographic records (authorship, source or year of publication, etc.), and citations themselves. A more in-depth discussion of the errors that can be found in Google Scholar has been recently published by Orduna-Malea, Martín-Martín, and Delgado López-Cózar [69].

**Table 21. Types of errors in Google Scholar**

| TYPE OF ERROR | DESCRIPTION |
|---|---|
| **Search-related** | Those related to the process of searching information |
| **Parsing-related** | related to the process of identifying and extracting bibliographic information about documents from websites or full-texts (including cited references) |
| **Matching** | Those related to the process of identifying different versions of a same document in order to remove duplicates. |
| **Source-related** | which affect the links that lead to the source in which the document has been found |

## 2.4.2 Google Scholar limitations

After describing the most common type of errors in the database, this section describes the main limitations for the use of GS as a source for bibliometric studies and research performance evaluation. To this end, we have prepared three descriptive sheets listing the limitations of Google Scholar Search, Google Scholar Citations, and Google Scholar Metrics, because although some limitations are present in all three products, some of them are particular to only a few of them.

Each descriptive sheet is contains several sections: coverage, search and results interface, quality of the data, and data reuse and exporting capabilities. The information included has been taken from the own authors' observations and empirical tests. Due to the long extension of the sheets, the sheet that describes the limitations of Google Scholar can be found below in Table 22, and the tables that list the limitations of Google Scholar Citations and Google Scholar Metrics can be found in the supplementary material [35] (see Appendix VI).

**Table 22. Google Scholar descriptive sheet**

| GOOGLE SCHOLAR |
|---|
| **COVERAGE** |
| Lack of transparency in its coverage*: <ul><li>There isn't a public master list of the sources Google Scholar indexes (publishers, repositories, catalogues, bibliographic databases and repertoires, aggregators…).</li><li>There isn't a public master list of journals indexed in the platform.</li></ul> |
| Non-scientific and non-academic documents are also occasionally covered: course syllabi, library guides, tweets… |
| There is no accurate method to estimate the size of Google Scholar. |
| Data is not stable. Google Scholar is dynamic and reflects the state of the academic web at a certain moment in time… The irregularity and unpredictability of Google Scholar's indexing speed may bias some bibliometric analyses if it is not taken under consideration. |
| Full text files that exceed 5MB are findable on Google Scholar, but their full text won't be indexed (cited references won't be analysed). |
| Easy to manipulate: anyone can get a fully or partially fabricated document indexed in GS by uploading it to a university domain or public academic repository. |
| **SEARCH AND RESULTS INTERFACE** |
| The advanced search form is limited to four search dimensions: keywords (with assisted Boolean operators, and the possibility to search only in the title of the document, or anywhere in the article), authors, source of publication (journal, conference…), and year of publication. |
| The number of records displayed in each results page is 10 by default. It can be increased to 20 in the settings page. In the past, however, it was possible to increase this number up to 100. |
| Only the first 1000 results for any query can be displayed. Similarly, even if a document has received more than 1000 citations, only the first 1000 can be displayed when clicking the "Cited by" link. |
| Results can only be sorted by relevance or by date of publication: <ul><li>Relevance: it is the default method. Although the specific parameters that are taking into consideration for this sorting method haven't been publicly disclosed, it has been found that the number of citations received by documents, as well as the language of the document in relation to the user's preferred language, both weight heavily in the relevance sorting algorithm.</li><li>Publication date: limits the search to documents published in the current year.</li></ul> |
| There are only three result filtering options once a search has been made: <ul><li>By document type, limited to three categories: case laws, patents, and articles. The latter category includes journal articles, books or book chapters, conference proceedings, technical reports, theses…</li><li>By type of record: users are given the option to remove cited references (documents GS has only been able to find in the reference lists of other documents) from the search results. By default, cited references are included in the search results.</li><li>By year: it is possible to limit results to documents published in a given year, or a range of years.</li></ul> |
| It does not offer any features to analyze results or compute bibliometric indicators. |
| **QUALITY OF THE DATA** |

| | |
|---|---|
| In each search result, **authors** are displayed in the second line: below the title and next to the source of publication (usually a journal), the date of publication, and the name of the publisher or web domain where the document has been found. The space allocated to the author data in this line is limited (usually between 30 and 40 characters), and therefore only the first three or four authors can be displayed, depending on the length of their names. In this line, authors are mentioned only by their first and second name initials and their surname. If these authors have created a public GSC profile and verified it with their institutional email, their name will contain a link to their public GSC profile.<br><br>For more complete author data, users can click the Cite button, and there, export the record to BibTeX (or other reference manager format). The BibTeX record will display the full name of up to 10 authors of the document. | |
| No data regarding **institutional affiliation** of the authors of the documents is available (institution, country). Therefore, it is not possible to carry out studies on geographic and institutional production and collaboration. | |
| There is no information available about the **language** in which documents are written, even though they must internally have this information, because users can choose to limit results to documents written in one or more of the following languages : Simplified Chinese, Traditional Chinese, Dutch, English, French, German, Italian, Japanese, Korean, Polish, Portuguese, Spanish, and Turkish. | |
| The **typology of each document** is not clear (book, journal article, conference communication, thesis, report…). Only books are marked as such, usually when they have been found on Google Books. | |
| Not all documents have an **abstract** | |
| **Author-supplied keywords** are not available. The same happens with the descriptors used by the databases where the records are found. | |
| The **list of cited references** in each article is not available either (even though they definitely have that information, since they need it to compute citations), making it difficult to carry out studies that require cited reference analysis. | |
| Errors in the parsing routine can lead to numerous problems. There still isn't a conclusive study about the type and degree of occurrence of these errors, but among them we can find:<br>• Poor bibliographic description of documents: incorrect or missing titles, authors, source of publication, date of publication…<br>• Duplicate records, when GS isn't able to match two or more records that actually refer to the same document. This also can lead to split citation counts, since some of the citations will be attributed to one of the versions of the document, and some to the other versions. | |
| Google Scholar doesn't rely in any kind of controlled vocabulary for author names, journals, publishers, institutions… that facilitates the identification of the different name variants for these entities. | |
| These last two limitations make it more difficult to carry out large scale studies using GS data, since the data would have to go through important cleaning and normalization processes prior to the analysis. | |
| **DATA REUSE AND EXPORTING CAPABILITIES** | |
| Users can copy citations for single records, in a variety of formats, by clicking in the "Cite" button below every record. | |

| |
|---|
| Records can be exported to reference managers manually one by one, also by clicking on the "Cite" button. The available formats are BibTeX, EndNote, RefMan, and RefWorks. Alternatively, users can also save records to the *My library* feature by clicking the "Save" button (for which it's necessary to be logged in to a gmail account). *My library* allows users to export up to 20 records in one go, in BibTeX, EndNote, RefMan, or CSV format. The abstract is never included as part of the exported records. |
| Google Scholar doesn't offer, nor plans to offer (at least in the near future) any kind of public API (Application Programming Interface) to enable users to access and export data from Google Scholar in bulk. |
| A strict CAPTCHA system is in place to discourage users from making too many queries to the platform too quickly. Users (or bots) that go over a certain (undisclosed) number of queries in a certain time are asked to solve a CAPTCHA of some sort before being able to continue their searches. Sometimes, if the system detects too many searches have been made from the same IP, that IP can get blocked temporarily. |

\* This limitation also affects Google Scholar Citations and Google Scholar Metrics

**2.5 The expanded academic world of Google Scholar**

The results of the analysis of the size, coverage, growth rate, and speed of indexing of GS fully justify considering this platform as a big data bibliographic tool (Figure 10).

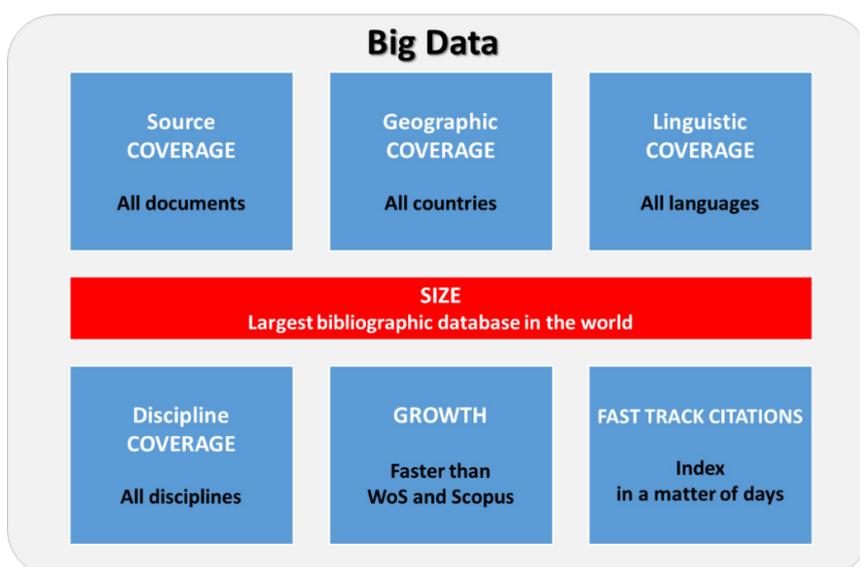

**Figure 10. Google Scholar, a big data bibliometric tool**

The empirical evidence described throughout this chapter allow us to affirm that GS is an all-inclusive tool, capable of bringing together not only the scientific world *stricto sensu* (that which is represented by WoS and Scopus as well), but the entire academic and professional world in a broad sense, thus providing a much broader picture of academic activity [36]. Its coverage is the most well-balanced of the commonly used multidisciplinary databases in terms of countries of publication, languages (no English bias), and document typologies (not only scientific articles), something crucial when analyzing fields where it is common to use other channels of communication other than journal articles published in English, such as disciplines in the Arts, Humanities, Social Sciences, and Engineering. In this sense, Figure 10 tries to convey the idea that GS covers all knowledge territories and all communities (scientific, educational, and professional), while WoS and Scopus only deal with scientific knowledge in the strict sense, and its communities.

The most important change over the previous paradigm, however, is Google Scholar's inclusion policy. WoS and Scopus have very exclusive and restrictive source inclusion policies. These sources are usually journals, which also place the prospective manuscripts researchers send to them under a rigorous evaluation processes (peer review). Opposite to this traditional model, GS works in a completely automated manner, without exercising any kind of selection process based on quality. Curated content from traditional journals and studies that haven't gone through any kind of screening both coexist. GS automatically crawls, finds, and indexes any document which follows an academic structure and is hosted in an academic domain, even if it hasn't undergone any external quality control and is there only by decision of its authors. This breaks completely from the traditional controls to which all academic content had to be subjected prior to its public dissemination (peer review). For better of for worse, this is the distinguishing feature of GS: its ability to bring together reviewed and non-reviewed content; scientific and academic content (Figure 11).

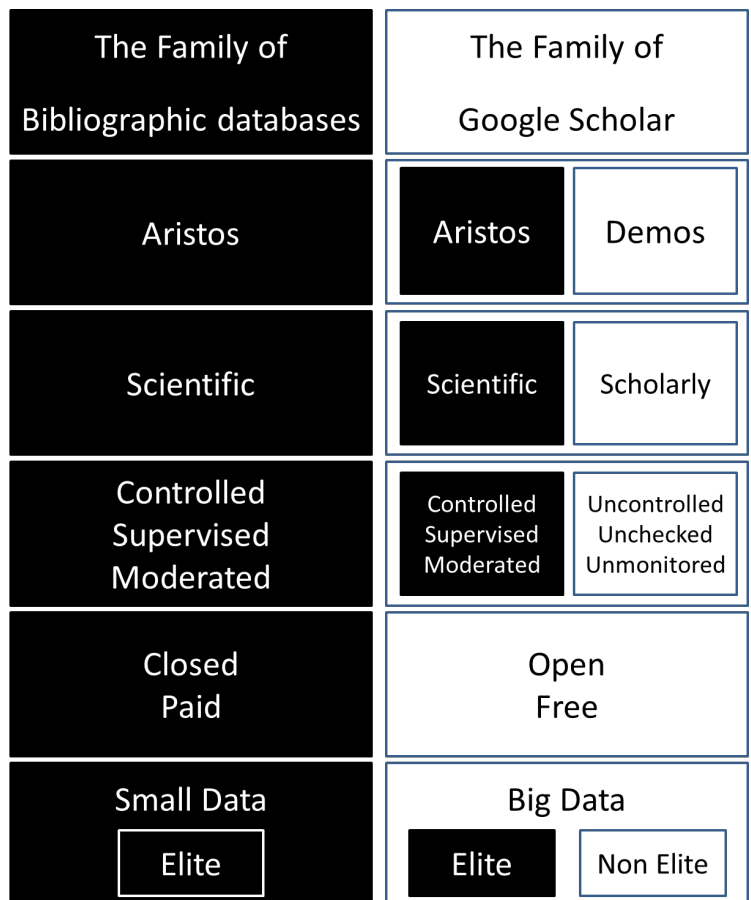

**Figure 11. Google Scholar versus traditional bibliographic databases**

It should be pointed out that one of the main features shown in Figure 11 is Google Scholar's nature as a receptacle. All the prestigious publications covered by WoS and Scopus are also covered by GS. When we observe the sources from which GS feeds, we can be sure, based on the empirical evidence on its size and coverage presented in the previous sections, that all the content covered by WoS and Scopus is also covered by GS. But of course, apart from the scientific elite, GS covers many other sources. We don't dispute that some of them may be of a lower quality, but other are of the same quality, if not higher, especially in certain disciplines (mainly doctoral theses, conference articles, working papers, and books).

Although it is true that GS covers sources that haven't gone through a validation process like peer review (keynote conferences, syllabi, book reviews, technical reports…) these sources provide evidence of other kinds of impact beside the scientific impact, and could help put in a new light the work of researchers whose work is relevant to these communities and not the ones who publish in traditional databases.

Of course, the mixture of all these source documents (especially when considering all of them are considered for computing citations) has been the object of important discussions in the bibliometric community [4]. To date, the main method to validate measurements made with data from GS has been to calculate their correlation with other well-established indicators. Many studies, which have been recently compiled by Thelwall y Kousha [70], have analysed correlations between the number of citations according to GS and other databases (mainly WoS and Scopus), either at the level of journals or at the level of authors, as a way to evaluate its suitability as a source of data bibliometric studies.

The supplementary material [35] offers a revised and updated list of these studies (see Appendixes II, III, and IV). Studies are grouped according to their unit of analysis (authors, citations, and average h-index). Although caution is advised for interpreting these results, because the natures of the samples are very different in terms of their size, analysed timeframes, and time when the studies were carried out, we can extract the following observations:

- The average correlation between GS and WoS (from 51 observations) is 0.76, and 0.81 when GS is compared to Scopus (28 observations).
- Out of all the studies, only in 2 the correlations found are below 0.50 (0.39 and 0.43 in Scopus, and 0.48 and 0.50 in WoS). All of them refer to correlations in the Humanities and Social Sciences. On the other hand, there are numerous studies where the correlations found are above 0.9 (1 with GS/WoS comparisons and 7 with GS/Scopus comparisons).
- The highest correlations are found among STEM fields, and the lowest ones are usually found for fields in the Humanities and Social Sciences.
- The differences between GS and the other databases seem to increase in the more recent studies, which might indicate an even broader coverage in GS respect to the other databases in recent years.
- No significative differences are found among studies with different units of analysis.

It also seems that multidisciplinary studies of international scope, and with very high sample sizes, achieve very high correlations, but that these become moderate by restricting the focus and emerging the intrinsic properties of each discipline. However, it should be borne in mind that correlations may have been calculated using different techniques (Pearson, Spearman, etc.), although values are reported independently, which may have a slightly influence on results.

To date, the largest sample studied for these purposes is the one used by Martin-Martin et al [71], who used a sample of 64,000 highly cited documents in Google Scholar, of which 51% were also covered by WoS and had at least one citation. Most of the documents covered by both databases were journal articles, and the rest were monographs, theses, and conference articles. The Spearman correlation found for number of citations received by documents covered by both databases was $R^2=0.73$.

For this chapter, we decided to replicate the previous study, using the sample of 861,843 highly-cited documents in 13 languages (see Appendix V for further details about this dataset) [35; 56] we used in section 1.3.2.5. Out of these documents, 69,279 (8%) of them were covered by WoS and had received at least one citation. The Spearman correlation between the number of citations according to GS and according to WoS for these documents was $R_s=0.9$. Figure 12 presents a scatterplot based on these data.

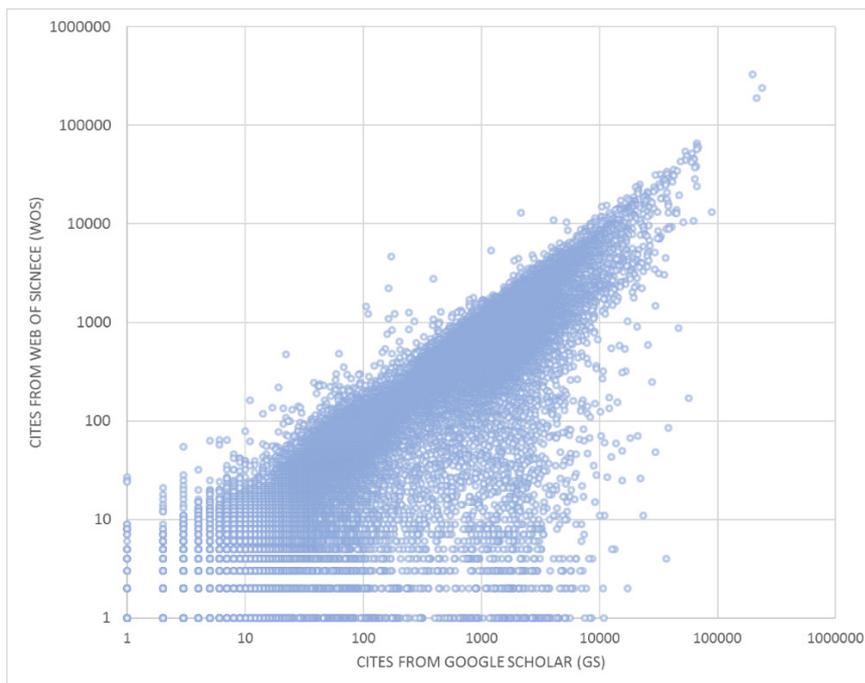

**Figure 12. Scatterplot of citations to highly cited documents according to Google Scholar and WoS (N=69,279)**

As Figure 12 shows, the correlation between the number of citations these documents have received according to the two databases is evident. Additionally, an important number of observations seem to have received many more citations from GS than from WoS. This means that, even though the correlation is very high in general terms, GS is usually able to find many more citations than WoS for the same documents.

**2.6 Final remarks**

Google Scholar is a prodigious mine of academic information that covers all fields of knowledge. Thanks to GS we now have access to previously unexplored territories of knowledge which, even if only roughly, are allowing us to form a broader mental picture of academic activity. This platform sheds light where previously there was only darkness. Google Scholar can help to definitely open the academic Pandora's box [4]. Opening this box will bring to light document typologies (especially monograpgs, theses, reports, conference communications, and book chapters) from the scientific core and periphery, which were previously invisible and unaccessible. The bibliographic features of these documents will probably change certain axioms and prejudices of academic evaluation.

Previously undervalued researchers, clearly harmed (or outright forgotten) by the policies of traditional databases, will come to light, because there will be evidende of the impact of their work. Conversely, it will also confirm the poor performance of other researchers, until now protected by the lack of proper tools to evaluate them.

Perhaps all this will finally lead to a redesign of certain assessment and promotion systems, funding programmes, and even research policies and structures. The report *The Metric Tide* [72] already gives a glimpse of this changing trend from an institutional position, and not merely as an intellectual exercise advocated by a few and confined to research publications, with varying degrees of scientific impact, but without an actual practical impact.

That said, we cannot belittle the limitations of GS for bibliometric analyses. To begin with, the search a exporting limitations (a maximum of 1,000 results per query, and no easy way to export them). These obstacles are a hindrance when massive amounts of data are necessary for an analysis. The lack of an API (Application Programming Interface) in Google Scholar forces us to use third party applications (like Harzing's Publish or Perish) or download results manually. This results in very slow and costly data collection processes, which must be followed by a thorough cleaning of the raw data [62; 73; 74].

Additionally, GS doesn't provide vital information in its records, like the institutional affiliation of the authors, the language, and the document types. Not to mention the difficulties that normalizing the bibliographic information collected from so many varied sources entails. On the other hand, we don't believe that the errors from which some records suffer are a major obstacle to use GS for bibliometric purposes. In a big data tool such as this, these errors are diluted and have no consequence on the big picture. These errors rarely affect individuals, journals, or other aggregates, as referred by some studies already [75].

At any rate, the more dangerous limitations do not have to do with the methodological and technical problems previously commented, but with the obscurity of the system, and, most of all, with the possibility of publication or citation manipulation, caused by the lack of quality control in the indexation of documents, which has been empirically proven by various studies [19; 76; 77]. Likewise, one of the main criticism that is directed at GS is the lack of transparency, both regarding its coverage (what sources it indexes) and updating mechanisms, and regarding its algorithm for ranking results after a query.

Lastly, we wish that this study helps readers understand the inner workings of Google Scholar, and become aware of its enormous potential. We always try to offer empirical evidence of its strengths (*contra data non argumenta*), without forgetting about the important dangers that its abuse could lead to. We hope to make the scientific community question assumed truths of an academic world of which only the tip of the iceberg has been visible until now. Let's explore its depths, let's observe and describe these new landscapes, and then let's decide if we'd rather remain on the surface.